\documentclass[3p]{elsarticle}

\usepackage{lineno}
\usepackage[hyphens]{url}
\usepackage{hyperref}
\hypersetup{breaklinks=true}
\modulolinenumbers[5]

\usepackage[utf8]{inputenc}

\usepackage{gensymb}
\usepackage{tabularx} 
\usepackage{amsmath}  
\usepackage{graphicx}
\usepackage{caption}
\usepackage{subcaption}
\usepackage{mathtools}
\usepackage{booktabs,arydshln}
\usepackage{adjustbox}

\usepackage[dvipsnames]{xcolor}
\definecolor{fgreen}{rgb}{0.13, 0.55, 0.13}
\journal{Journal of Solar Energy Materials \& Solar Cells}

\begin{document}

\begin{frontmatter}

\title{Monolithic Thin-Film Chalcogenide-Silicon Tandem Solar Cells Enabled by a Diffusion Barrier}


\author[mymainaddress]{Alireza Hajijafarassar\corref{mycorrespondingauthor}\fnref{myfootnote}}
\ead{alhaj@dtu.dk}

\author[mysecondaryaddress]{Filipe Martinho\corref{mycorrespondingauthor}\fnref{myfootnote}}
\ead{filim@fotonik.dtu.dk}
\fntext[myfootnote]{These authors contributed equally.}
\cortext[mycorrespondingauthor]{Corresponding authors}

\author[mythirdaddress]{Fredrik Stulen}

\author[mythirdaddress]{Sigbjørn Grini}

\author[mymainaddress]{Simón López-Mariño}

\author[mysecondaryaddress]{Moises Espíndola-Rodríguez}

\author[myforthaddress]{Max Döbeli}

\author[mysecondaryaddress]{Stela Canulescu}

\author[mymainaddress]{Eugen Stamate}

\author[mysecondaryaddress]{Mungunshagai Gansukh}

\author[mysecondaryaddress]{Sara Engberg}

\author[myfifthaddress]{Andrea Crovetto}

\author[mythirdaddress]{Lasse Vines}

\author[mysecondaryaddress]{Jørgen Schou}

\author[mymainaddress]{Ole Hansen}

\address[mymainaddress]{DTU Nanolab, Technical University of Denmark, DK-2800 Kgs. Lyngby, Denmark}

\address[mysecondaryaddress]{Department of Photonics Engineering, Technical University of Denmark, DK-4000 Roskilde, Denmark}

\address[mythirdaddress]{Department of Physics, University of Oslo, 0371 Oslo, Norway}

\address[myforthaddress]{Ion Beam Physics, ETH Zurich, CH-8093 Zurich, Switzerland}

\address[myfifthaddress]{DTU Physics, Technical University of Denmark, DK-2800 Kgs. Lyngby, Denmark}

\begin{abstract}
Following the recent success of monolithically integrated Perovskite/Si tandem solar cells, great interest has been raised in searching for alternative wide bandgap top-cell materials with prospects of a fully earth-abundant, stable and efficient tandem solar cell. Thin film chalcogenides (TFCs) such as the Cu\textsubscript{2}ZnSnS\textsubscript{4} (CZTS) could be suitable top-cell materials. However, TFCs have the disadvantage that generally at least one high temperature step ($>$ 500 \degree C) is needed during the synthesis, which could contaminate the Si bottom cell. Here, we systematically investigate the monolithic integration of CZTS on a Si bottom solar cell. A thermally resilient double-sided Tunnel Oxide Passivated Contact (TOPCon) structure is used as bottom cell. A thin ($<$ 25 nm) TiN layer between the top and bottom cells, doubles as diffusion barrier and recombination layer. We show that TiN successfully mitigates in-diffusion of CZTS elements into the c-Si bulk during the high temperature sulfurization process, and find no evidence of electrically active deep Si bulk defects in samples protected by just 10 nm TiN. Post-process minority carrier lifetime in Si exceeded 1.5 ms, i.e., a promising implied open-circuit voltage (i-\textit{V}\textsubscript{oc}) of 715 mV after the high temperature sulfurization. Based on these results, we demonstrate a first proof-of-concept two-terminal CZTS/Si tandem device with an efficiency of 1.1\% and a V\textsubscript{oc} of 900 mV. A general implication of this study is that the growth of complex semiconductors on Si using high temperature steps is technically feasible, and can potentially lead to efficient monolithically integrated two-terminal tandem solar cells.
\end{abstract}

\begin{keyword}
Tandem\sep Photovoltaic\sep Silicon\sep TOPCon\sep CZTS\sep Titanium Nitride
\MSC[2019] 00-01\sep  99-00
\end{keyword}

\end{frontmatter}


\section{Introduction}
The current global uptake of photovoltaic (PV)-based solar energy has been enabled by the remarkable developments in crystalline Silicon (c-Si) solar cell technologies, both in terms of module efficiencies and costs, with market shares consistently around 90\% for decades – a figure which is expected to remain unchanged in the near future \cite{osti_1344202, VDMA2018, MarkusFischer2019}.  However, as the Si cell efficiency approaches the Shockley-Queisser (SQ) single-junction limit \cite{doi:10.1063/1.1736034}, further cell improvements are now only incremental, and the focus is instead on systems cost reduction and raw material utilization \cite{VDMA2018, MarkusFischer2019}.

Multi-junction solar cells can achieve higher efficiencies than the single-junction SQ limit, with AM 1.5 limits of around 45\% and 50.5\% for double (also called tandem) and triple-junction solar cells, respectively \cite{Green1982, Green2014}. However, to transition the global PV market from a single- to a multi-junction solar cell technology, the following conditions must be met: 1) The efficiency improvements should not sacrifice cost competitiveness, namely in terms of the Levelized Cost of Electricity (LCOE); 2) The raw materials used should be abundant, inexpensive and non-toxic; 3) Each individual junction, as well as the full device, must be stable and have a lifetime of decades {\cite{Yu2018, Bae2019}}. 

Various multi-junction cell configurations have been proposed and demonstrated experimentally, particularly with Si and III-V semiconductors, reaching efficiencies of 32.8\%  and 37.9\%  for tandem and triple-junction cells, respectively \cite{Yu2016, Essig2017, doi:10.1002/pip.3102, SharpCorporation2013}. In space applications, multi-junction III-V solar cells have been used almost exclusively since the late 1990s, but with costs not competitive with the single-junction c-Si technology for terrestrial applications \cite{ILES20011, 5950320}. Out of all the possible multi-junction configurations, a monolithically integrated two-terminal (MI-2T) tandem device is considered \textit{a priori} to be the most feasible for cost-competitive, large-scale applications, since it retains the module design simplicity of single-junction technologies and minimizes the total number of processing steps. Despite all of its potential advantages, MI-2T tandem devices are challenging to achieve in practice because every processing step has to be compatible and the properties of the preceding interface and layers should not be compromised \cite{C6ME00041J}.

c-Si is an excellent partner in a tandem solar cell for the same reasons that gave it a dominant position in the PV market. Its band gap of 1.12 eV is near ideal for a MI-2T tandem – when used together with an absorber with a bandgap of 1.72 eV, a theoretical maximum efficiency of close to 43\%  can be achieved \cite{Green2014, Yu2016}. Recently, a lot of interest has been raised after a series of MI-2T Perovskite/Si tandem devices achieved efficiencies above 25\%, with the current record set at 28\%  \cite{Sahli2018, OxfordPV2018, NREL2019}, a value higher than that of the best Si solar cell.

Thin film chalcogenides (TFCs) such as CdTe, CuIn\textsubscript{x}Ga\textsubscript{1-x}(S\textsubscript{y}Se\textsubscript{1-y})\textsubscript{2} (CIGSSe), Cu\textsubscript{2}ZnSn(S\textsubscript{x}Se\textsubscript{1-x})\textsubscript{4} (CZTSSe) and their respective solid solutions and cationic substitutions could be suitable alternatives to Perovskites due to their increasing single-junction solar cell efficiencies, competitive production costs and superior stability. Indeed, a 16.8\%  Cd\textsubscript{1-x}Zn\textsubscript{x}Te/Si tandem cell has been demonstrated using low temperature molecular beam epitaxy (MBE) \cite{doi:10.1063/1.3582902}. However, in most cases, TFCs have the disadvantage that a high temperature step $>$ 500  \degree C ) is needed, contrary to Perovskites which can be processed at low temperatures ($<$ 200 \degree C) \cite{C6ME00041J, Sahli2018}. Recently, a promising monolithic Si/CGSe tandem cell with an efficiency of 10\% has been reported \cite{Jeong2017}, where the CGSe layer was produced by co-evaporation and high-temperature annealing. The authors report that the bottom Si cell J-V curve was not degraded during the top cell processing, however no further details were provided regarding the characterization of the bottom Si cell, and it is unclear what would be the resilience of the bottom cell for different processing parameters, or different deposition methods. Hence, to the best of our knowledge, the implications of high temperature processing on the feasibility of a MI-2T TFC/Si tandem device remain relatively unknown and have not yet been directly assessed experimentally.

In this work, we discuss the challenges of producing TFC/Si MI-2T tandem devices, using the sulfide kesterite Cu\textsubscript{2}ZnSnS\textsubscript{4} (CZTS), an earth abundant and environmentally friendly representative of the TFC group. In particular, we assess the contamination and degradation of a Tunnel Oxide Passivated Contact (TOPCon) Si bottom cell during the CZTS processing steps. We test the introduction of a thin titanium nitride (TiN) diffusion barrier layer between the Si and CZTS structures and use the results to evaluate the process compatibility between CZTS and Si. We show that compatibility can be achieved, and report on a first proof of concept CZTS/Si tandem solar cell with an efficiency of 1.1\%  and a \textit{V}\textsubscript{oc} of 900 mV, a value higher than that of each respective individual reference cell. Moreover, we suggest strategies for future device improvement.

\subsection{The Top Cell: CZTS}
The kesterite sulfide-selenide CZTSSe attracted interest as an all earth abundant alternative to CIGSSe consisting of non-toxic elements (in particular the sulfide CZTS), achieving solar cell efficiencies above 10\%  using industrially upscalable methods such as sputtering \cite{Yan2018}. Sulfide CZTS, in particular, has some features which suggest that it could be a promising tandem partner for Si. Through different solid solutions and cationic substitutions, the bandgap of kesterites can be tuned – for instance, through Ge or Ag incorporation the bandgap of sulfide CZTS can be increased from the nominal 1.5 eV to about 2.1 eV, an ideal range for tandem applications \cite{OUESLATI2019315, GARCIALLAMAS2016147, UMEHARA2016713, doi:10.1002/pssc.201400343, doi:10.1021/acs.chemmater.8b00677, doi:10.1063/1.4863951}.  Moreover, CZTS and Si are closely lattice-matched, with an a-axis lattice mismatch of less than $\pm$ 0.1\% \cite{doi:10.1002/9781119052814.ch2, doi:10.1002/9781118437865.ch3}. This means that heteroepitaxial growth of CZTS on Si could be possible, and this has indeed been proven experimentally \cite{OISHI20081449, SHIN20149, doi:10.1063/1.4922992}. While this allows in principle for growing CZTS/Si tandem devices epitaxially (free of grain boundaries), epitaxial growth of CZTS on Si with the necessary tunnel junction structures has not been demonstrated yet.

So far, the TFC solar cells with the highest efficiencies, in particular in the case of CZTS, involved at least one high temperature step \cite{doi:10.1002/9781118437865.ch3} (with the notable exceptions of MBE \cite{doi:10.1063/1.3582902} and monograin technology \cite{MELLIKOV200965}). Herein, we argue that one of the biggest challenges towards a TFC/Si MI-2T tandem device could be a cross-contamination of the bottom Si cell with metallic elements such as Cu or chalcogens like S, during the high temperature step.

\subsection{The Bottom Silicon Cell: Tunnel Oxide Passivating Contacts (TOPCon)}

The tunnel oxide passivating contact (TOPCon) structure has played a key role in the recent silicon solar cell efficiency improvements \cite{Haase_2017, Feldmann2013, HAASE2018184, FELDMANN2014270, FELDMANN201446}. The structure consists of stacks of thin ($ \sim $ 1.2 -- 1.5 nm) SiO\textsubscript{2} layers (tunnelling oxide, TO) and highly doped (Phosphorous or Boron) polycrystalline silicon layers (PolySi) on both sides of a crystalline silicon (c-Si) wafer. This structure provides excellent surface passivation and carrier selectivity. Consequently, high implied \textit{V}\textsubscript{oc} of 750 mV and external \textit{V}\textsubscript{oc} of up to 739 mV have been achieved \cite{6960882}. In contrast to its aSi:H heterojunction counterpart, the TOPCon structure alone is resilient to high temperature annealing up to 900 \degree C, which is well above the typical annealing temperatures used in the synthesis of chalcogenide semiconductors and in other front and backend processes. Moreover, the simple one-dimensional current transport and full coverage of contacts at both sides allows for very low contact resistivity and thereby low fill-factor (FF) losses \cite{7747536}. A major drawback of a front PolySi contact in a single-junction device is the parasitic absorption losses in the blue wavelength region within the PolySi layer. As a result, a short-circuit current density (\textit{J}\textsubscript{sc}) loss of 0.5 mA/cm\textsuperscript{2} is expected for every 10 nm of PolySi in a single junction cell \cite{FELDMANN2017265}. However, this loss is not a limitation in a tandem configuration, where the high-energy photons are absorbed in the top cell. Thus, the double-sided TOPCon structure may be an ideal candidate for double-junction tandem solar cell.

\subsection{The Need for a Diffusion Barrier Layer}

When a nearly complete silicon solar cell is used as substrate for the growth of a TFC, there is a risk of contamination from metallic and chalcogen elements that should be thoroughly assessed. In this contribution, we study the case of co-sputtered CZTS precursors from Cu, ZnS and SnS targets on c-Si model substrates. During co-sputtering, the impinging energetic ions and neutrals can directly cause sputter damage, or contaminate the Si bulk by implantation. After co-sputtering, CZTS is formed by high temperature reactive annealing in a sulfur atmosphere. Here, the elements Cu, Zn, Sn and S (the latter both from the precursors and from the atmosphere) may diffuse into the Si bulk. We note that this high temperature step is of particular interest, since it is nearly ubiquitous in high-quality TFC fabrication, even in single-step processes (for instance co-evaporation of CIGS).

Copper contamination in silicon deserves special consideration as it is a common element of both the CIGS and the CZTS group of alloys and, most importantly, because it is one of the most common detrimental contaminants known in crystalline Si, as widely reported in the photovoltaic and integrated circuit industries \cite{osti_15000243,Istratov1998, Istratov01012002}. Copper has a high diffusivity in Si, and can diffuse through the entire thickness of a Si wafer at room temperature in a matter of hours, although the solid solubility is $<$ 10\textsuperscript{15} cm\textsuperscript{-3} at the relevant temperatures \cite{osti_15000243}. Cu exhibits a complex defect physics in Si, leading to point defects and complexes, decoration of extended defects, precipitation of copper silicides, out-diffusion to the surface and segregation phenomena. In particular, copper silicides have been shown to lead to mid-gap defect traps in Si and a high recombination activity \cite{Seibt2009}, detrimental in solar cells.

Although studied to a lesser extent than copper, the other elements of CZTS could also be harmful contaminants for a bottom Si cell. Zinc can introduce near-midgap defect levels in Si as shown in pure diffusion studies \cite{SZE1968599, Masuhr_1999, PhysRev.105.379}. Tin was studied in particular as a dopant to improve the radiation resistance of c-Si devices, but was also found to form midgap states in Si \cite{Larsen2001, PhysRevB.62.4535}. Finally, sulfur was studied notably in ``black silicon"  processing, where it was found that its incorporation creates deep bandgap states, which increase the infrared light absorption in Si, making it appear more ``black" \cite{SZE1968599, KOYAMA1978953, PhysRevB.70.205210}.

Here, we suggest that one possible way to prevent bottom cell contamination is using a diffusion barrier layer at the bottom-cell/top-cell interface. In general, a barrier layer must have properties such as mechanical stability, good adhesion, high temperature stability and low diffusivity for the contaminating elements. For tandem solar cell applications, it must also be electrically conductive and transparent in the near infrared region. To the best of our knowledge, only one published work directly addresses this problem, suggesting the use of ZnS as a barrier layer for the growth of CZTS/Si tandem cells \cite{7749871}. In this work, we propose titanium nitride (TiN) as a barrier layer at the CZTS/Si interface,  a novel concept for the monolithic integration of a top thin-film cell on a bottom Si cell. TiN has been extensively studied as a barrier layer for copper metallization in integrated circuits, although it is arguably not the most effective barrier known against Cu diffusion \cite{Istratov01012002, doi:10.1146/annurev.matsci.30.1.363, Lee2007}. TiN has been employed as a back contact modification and barrier against over-sulfurization (or over-selenization) in single-junction CZTSSe cells, and proved to be compatible with up to 9\%  efficiency devices \cite{Oueslati_2014, SCHNABEL2017290, ENGLUND201791, doi:10.1063/1.4740276, doi:10.1021/cm4015223}. Due to its poor transparency, the TiN thickness must be limited to only a few nm.

By contrast, in a MI-2T Perovskite/Si tandem solar cell, a Si-based tunnel junction or a simple interface recombination layer based on a transparent conductive oxide (TCO) can be used to achieve high performing devices \cite{C6ME00041J, Sahli2018}. This could also be a possibility if contamination-free growth of TFCs on Si can be proven. In this regard, it is noteworthy to mention that there are studies suggesting that some TCO substrates could be compatible with TFC growth conditions \cite{UMEHARA2016713, doi:10.1021/acssuschemeng.7b02797}.
\section{Materials and Methods}

A set of double side polished 100 mm diameter, 1 $\Omega$.cm, 350 $\mu$m thick,  (100) n-type Cz-Si wafers were used.

 The fabrication process of the TOPCon structure is as follows. After the wafers were cleaned in RCA1 (H\textsubscript{2}O\textsubscript{2}:NH\textsubscript{4}OH:5H\textsubscript{2}O) and RCA2 (H\textsubscript{2}O\textsubscript{2}:HCl:5H\textsubscript{2}O) mixtures, $\sim$ 1.2 nm of SiO\textsubscript{2} (tunnel oxide or TO) was grown by chemical oxidation in a 65 \%\textsubscript{wt} HNO\textsubscript{3} solution at 95 \degree C. Subsequently, $\sim$ 40 nm PolySi layers were deposited using Low Pressure Chemical Vapor Deposition (LPCVD) at 620\degree C, using SiH\textsubscript{4}, B\textsubscript{2}H\textsubscript{6}, or PH\textsubscript{3} as precursors for p+ or n+ PolySi layers, respectively. The samples were then annealed in N\textsubscript{2} at 850 \degree C for 20 min for further dopant diffusion and activation. All samples have a symmetrical passivation of TO/n+PolySi on both sides, except in two cases: for Deep Level Transient Spectroscopy (DLTS), this passivating stack was not used, and for the tandem solar cell fabrication, an asymmetrical passivation was used, with TO/n+PolySi on the front and TO/p+PolySi on the rear side. In the fabrication of tandem cells, a hydrogenation process was performed on the as-passivated bottom cell precursor wafer. For this purpose, a sacrificial $\sim$ 75 nm hydrogenated SiN (SiN:H) layer was deposited on both sides of the wafer using Plasma Enhanced Chemical Vapor Deposition (PECVD) at 300 \degree C. After a hydrogen drive-in process at 400 \degree C for 30 min in N\textsubscript{2} atmosphere, the SiN:H layers were stripped in a buffered HF solution. The benefits of this SiN hydrogenation process are similar to those achieved by annealing in forming gas \cite{Bae2017}. A few experiments used an alternative surface passivation with 40 nm Al\textsubscript{2}O\textsubscript{3}, deposited by Atomic Layer Deposition (ALD) using tetramethylammonia (TMA) and H\textsubscript{2}O as precursors.

TiN barrier layers ($<$ 25 nm) were deposited in a Picosun Plasma-Enhanced ALD (PEALD) system using TiCl\textsubscript{4} and NH\textsubscript{3} precursors at 500 \degree C. To improve the optical transparency of TiN, the ALD chamber was not passivated for nitride depositions in the tandem cell fabrication. As a result, a slightly higher amount of oxygen is present in the TiN layer used in the tandem cell. Metallic 100 nm Cu layers were sputtered on the TOPCon structure and annealed at 550  \degree C  in vacuum ($1\times10^{-6}$ mbar). CZTS precursors were co-sputtered from Cu, ZnS, and SnS targets, and annealed in a graphite box with a reactive N\textsubscript{2} atmosphere containing 50 mg of S pellets, at dwell temperatures of 525 -- 575 \degree C for 30 min, in order to form CZTS films with a thickness around 300 nm. This CZTS thickness was chosen based on optical simulations (not shown here) and photocurrent density (\textit{J}\textsubscript{sc}) results from single junction CZTS devices, in order to match the photocurrents of the two cells. Prior to lifetime, Secondary Ion Mass Spectrometry (SIMS), and DLTS measurements, the Cu, CZTS and TiN layers were removed (after the sulfurization/annealing step) in a mixture of H\textsubscript{2}O\textsubscript{2}:4H\textsubscript{2}SO\textsubscript{4} (piranha) and RCA1 solutions, followed by a dilute HF dip. For the Rutherford Backscattering Spectroscopy (RBS) measurements, only piranha was used.

The effective minority carrier lifetime ($\tau$\textsubscript{eff}) of Si was measured by the microwave detected photoconductance decay method ($\mu$-PCD) in steady-state configuration at 1-sun illumination using an MDP lifetime scanner from Freiberg Instruments. The reported lifetime values were obtained from maps of the whole wafer area with 1 cm edge-exclusion margin. The i-\textit{V}\textsubscript{oc} values were calculated based on the method described in \cite{doi:10.1002/aenm.201900439}.

The in-diffusion depth profile were measured by SIMS and RBS on selected samples. The SIMS depth profiles were obtained from a Cameca IMS-7f microprobe. A 10 keV O\textsubscript{2}\textsuperscript{+} primary beam was mainly utilized, and rastered over  $150\times150$  $\mu$m\textsuperscript{2}, and the positive ions were collected from a circular area with a diameter of 33  $\mu$m. For sulfur, however, a 5 keV Cs\textsuperscript{+} primary beam was employed, and clusters of \textsuperscript{32}S\textsuperscript{133}Cs were detected to minimize matrix effect and avoid mass interference. The quantification of Cu depth profiles was obtained by measuring an implanted reference sample, ensuring a  $\pm$ 10 \%  error in accuracy. The crater depths were measured by a Dektak 8 stylus profilometer, and a constant sputter erosion rate was assumed for the depth calculation. The RBS measurements were done using 2 MeV He ions and a silicon PIN diode detector under a 168$^{\circ}$  angle. The collected RBS data were analyzed and fitted using RUMP \cite{doi:10.1063/1.3340459}.

DLTS was used to characterize electrically active defects in the Si bulk. DLTS measurements were performed on circular Schottky diodes (1 mm diameter), where 50 nm thick Pd contacts were deposited by thermal evaporation. The backsides were coated with silver paste to form an ohmic contact. During the measurements the diodes were held at $-5$ V reverse bias and pulsed to 1 V, filling all majority traps within the depletion width of  $\sim$ 1  $\mu$m. The samples were cooled to 35 K by a closed-cycle cryostat and six rate windows (with lengths 2$^i$ $\times$ 10 ms, $i$ = 1, $\ldots$ , 6), were used to record the capacitance transients while heating to 300 K. The transients were multiplied by a lock-in weighting function for improved signal extraction. Further details on the method and setup are given in \cite{Liu2017}.

For the monolithic CZTS-Si tandem device, a 50 nm CdS layer, by chemical bath deposition, was used as the buffer to form the p-n heterojunction of the top cell, followed by a 50 nm intrinsic i-ZnO and a 350 nm Al-doped ZnO (AZO) as the TCO layer. Both window layers, i-ZnO and AZO, were deposited using reactive sputtering. A 500 nm Ag layer was thermally evaporated as the back contact. No front metal contacts were used for simplicity, as the active tandem cell areas were only 3$\times$3 mm\textsuperscript{2}. The full tandem solar cell was post-annealed on a hot plate in air at 250 \degree C for 15 min, in order to improve the properties of the CZTS/CdS heterojunction (see Figure S13).

The J-V characteristic curves of the solar cells were measured at near Standard Test Conditions (STC: 1000 W/m\textsuperscript{2}, AM 1.5 and 25 \degree C). A Newport class ABA steady state solar simulator was used. The irradiance was measured with a $2\times2$ cm\textsuperscript{2} Mono-Si reference cell from ReRa certified at STC by the Nijmegen PV measurement facility. The temperature was kept at 25 $\pm$ 3 \degree C as measured by a temperature probe on the contact plate. The acquisition was done with 2 ms between points, using a 4 wire measurement probe, from reverse to forward voltage. The external quantum efficiency (EQE) of the tandem cell was measured using a QEXL setup (PV Measurements) equipped with a grating monochromator, adjustable bias voltage, and a bias spectrum.

Room temperature photoluminescence (PL) measurements were done on complete cells with an excitation wavelength of 785 nm using a modified Renishaw Raman spectrometer equipped with a Si CCD detector, in confocal mode.

Scanning electron microscopy (SEM) images of the tandem cell structures were acquired using a Zeiss Merlin field emission electron microscope under a 5 kV acceleration voltage.

\section{Results and Discussion}

\subsection{Minority Carrier Lifetime Measurements on Si}\label{sec:31}

\begin{figure}
\centering
\begin{subfigure}[b]{.3\textwidth}
  \centering
  \includegraphics[width=\linewidth]{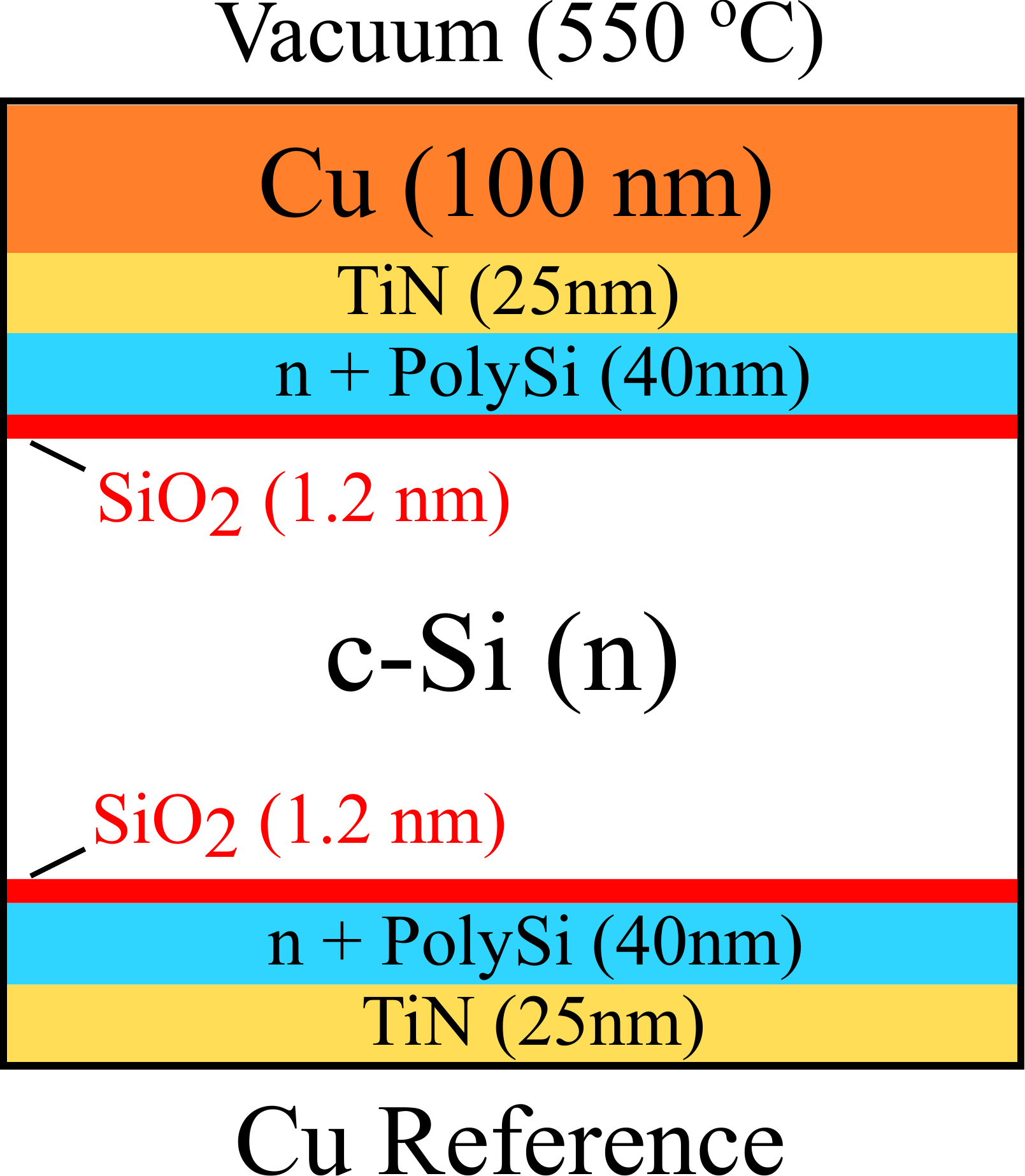}
  \caption{}
  \label{fig:1a}
\end{subfigure}
\begin{subfigure}[b]{.3\textwidth}
  \centering
  \includegraphics[width=\linewidth]{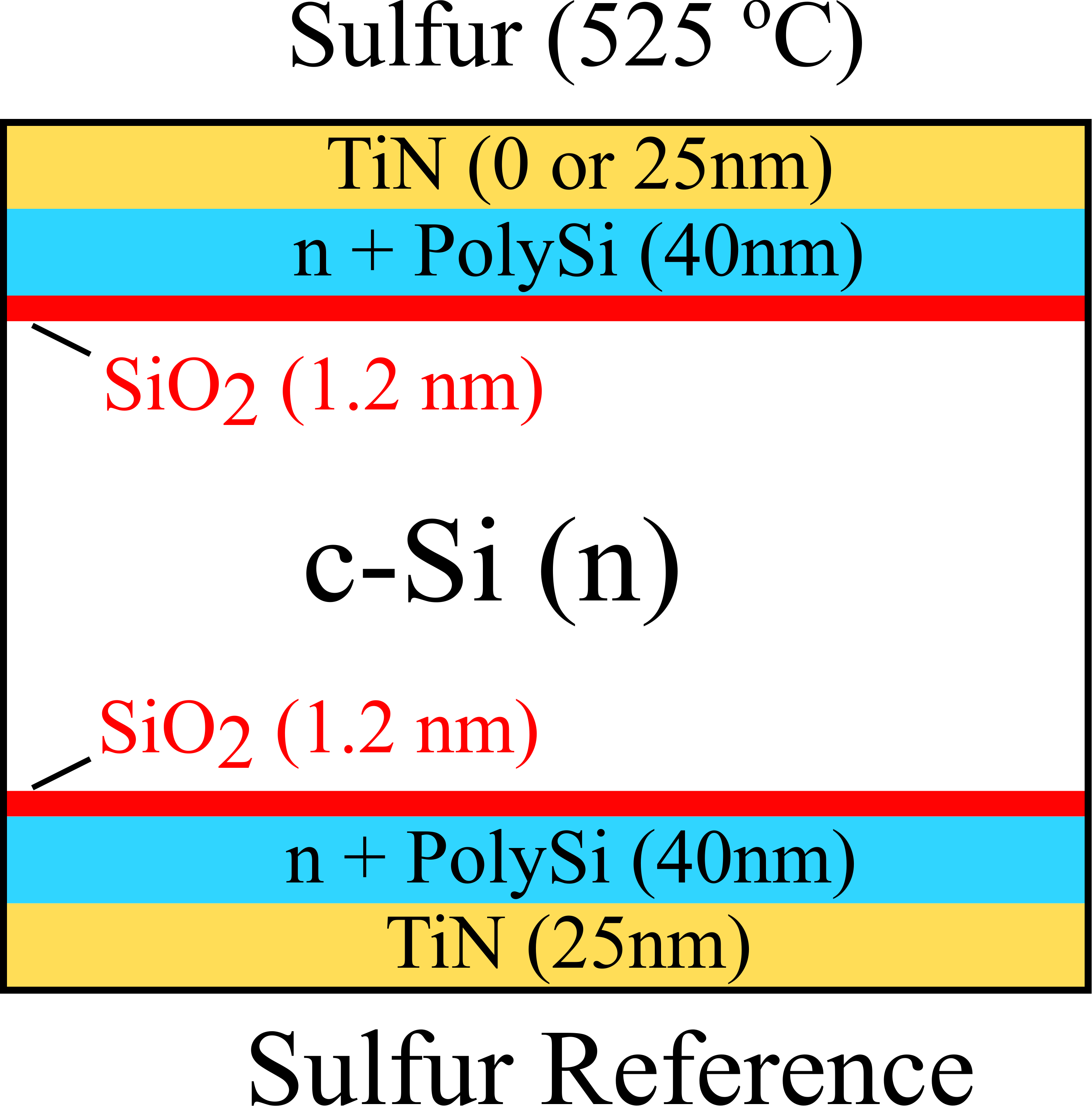}
  \caption{}
  \label{fig:1b}
\end{subfigure}
\begin{subfigure}[b]{.3\textwidth}
  \centering
  \includegraphics[width=\linewidth]{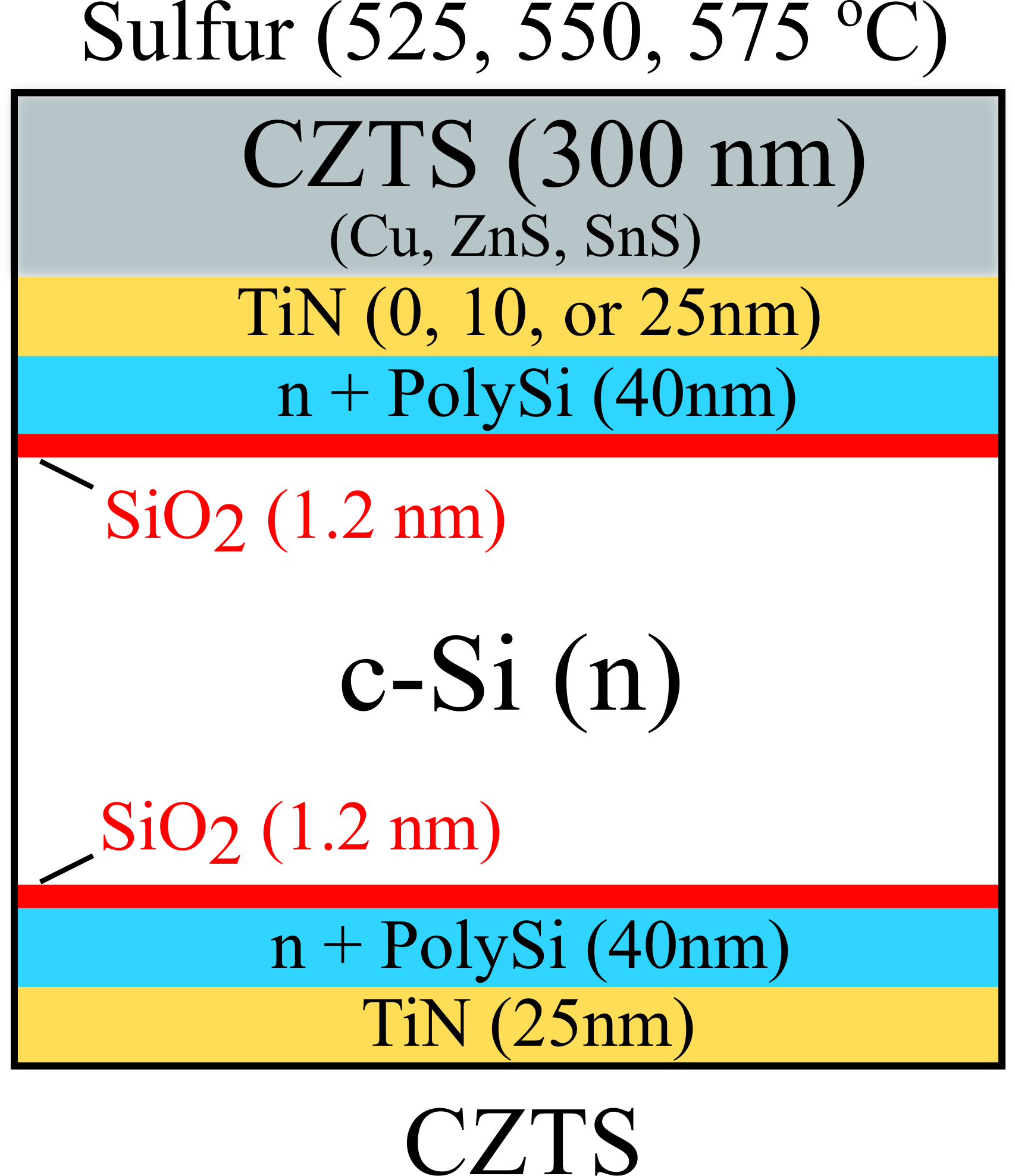}
  \caption{}
  \label{fig:1c}
\end{subfigure}
\caption{Cross-section scheme of the samples used for minority carrier lifetime measurements: a) \textit{Cu Reference}, b) \textit{Sulfur Reference}, and c) \textit{CZTS}}
\label{fig:1}
\end{figure}

\begin{figure}
  \centering
  \includegraphics[width=0.6\linewidth]{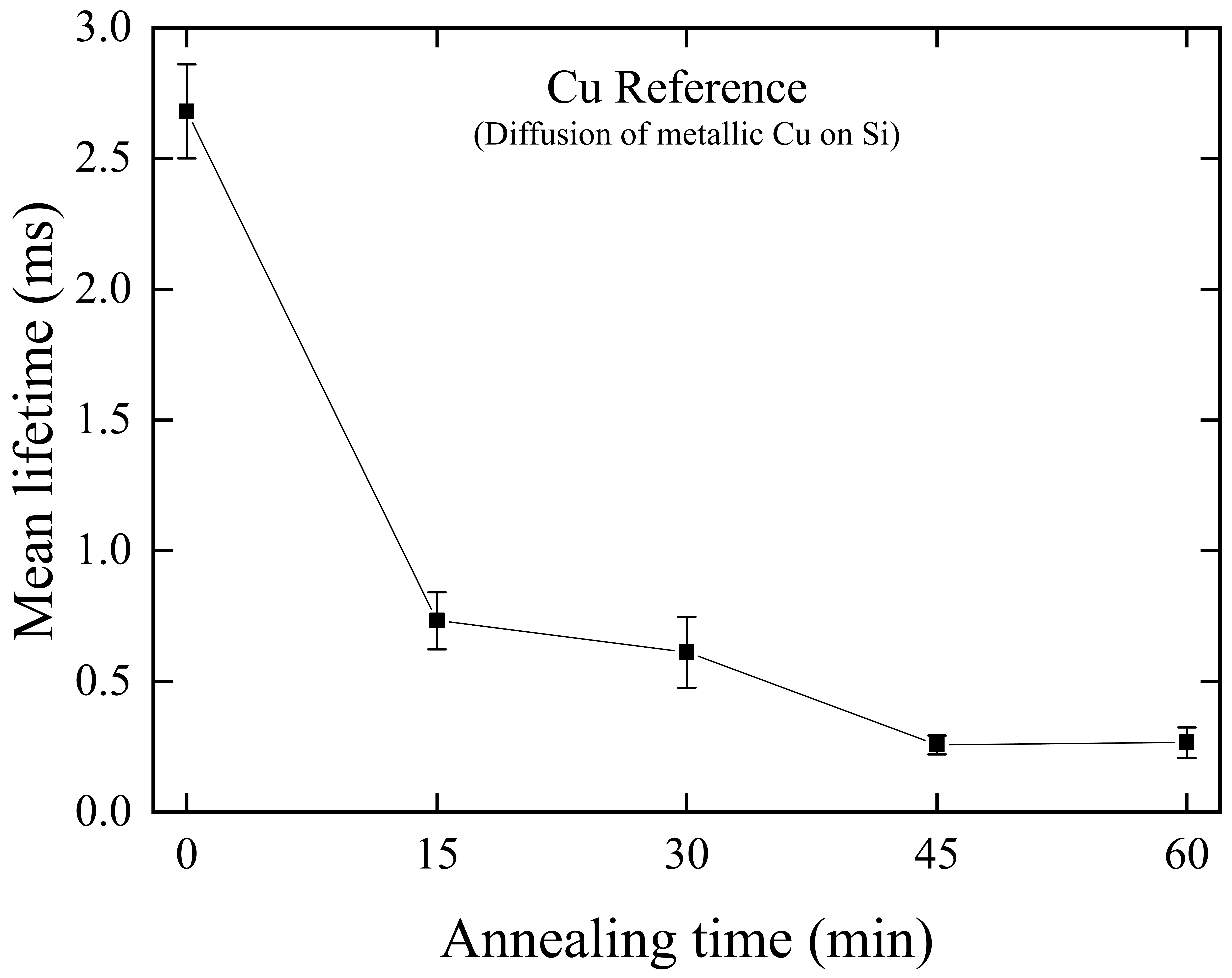}
\caption{Minority carrier lifetime evolution of the \textit{Cu Reference} sample with annealing time at 550 \degree C in vacuum.}
\label{fig:2}
\end{figure}

The minority carrier lifetime of Si is used as a figure of merit throughout the paper to evaluate the bottom cell after CZTS and tandem cell processing. For this purpose, 10 symmetrically passivated wafers with an as-passivated mean lifetime of 2.65 $ \pm $  0.52 ms were prepared. The uniform surface passivation quality across the wafer set ensures that the passivation and wafer qualities are not variables in the subsequent studies. More details on the passivation statistics are shown in the supplementary information (Figure S1).\  Three different set of samples were prepared, as listed in Table \ref{tab:1} and illustrated in Figure \ref{fig:1}. All the samples have a 25 nm TiN layer on the backside, to eliminate any unwanted contamination from that side during the different processing steps.
\begin{table*}[ht]
    \renewcommand{\arraystretch}{1}
    \begin{center}
    \caption{Overview of the different samples used for minority carrier lifetime measurements. Note: all the samples have 25 nm TiN on the backside.}
    \begin{tabular}{llll}
    \toprule
      \textbf{Sample} & \textbf{TiN Thickness (nm)} & \textbf{Annealing atm./T \degree C} &\textbf{Purpose} \\ \midrule
      \textit{Cu Reference} &25 &Vacuum/550 &Compare metallic Cu to CZTS \\
       \textit{Sulfur Reference} &0, 10 &Sulfur/525 &Isolate the effect of S \\
        \textit{CZTS} &0, 10, 25 &Sulfur/525, 550, 575 &Integration of CZTS on Si \\
      \bottomrule
    \end{tabular}
    \label{tab:1}
  \end{center}
\end{table*}
In Figure \ref{fig:2}, the Si minority carrier lifetime of the \textit{Cu Reference} sample is shown as a function of the annealing time. In this case, the TiN barrier layer fails after a 15 min annealing at 550 \degree C , with a 73\%  loss of lifetime. The lifetime is further degraded with increasing annealing time. These results indicate that, for this temperature range, the 25 nm TiN barrier layer provided insufficient protection of the sample against Cu diffusion.

\begin{figure}
\centering
\begin{subfigure}[b]{0.6\textwidth}
  \centering
  \includegraphics[width=\linewidth]{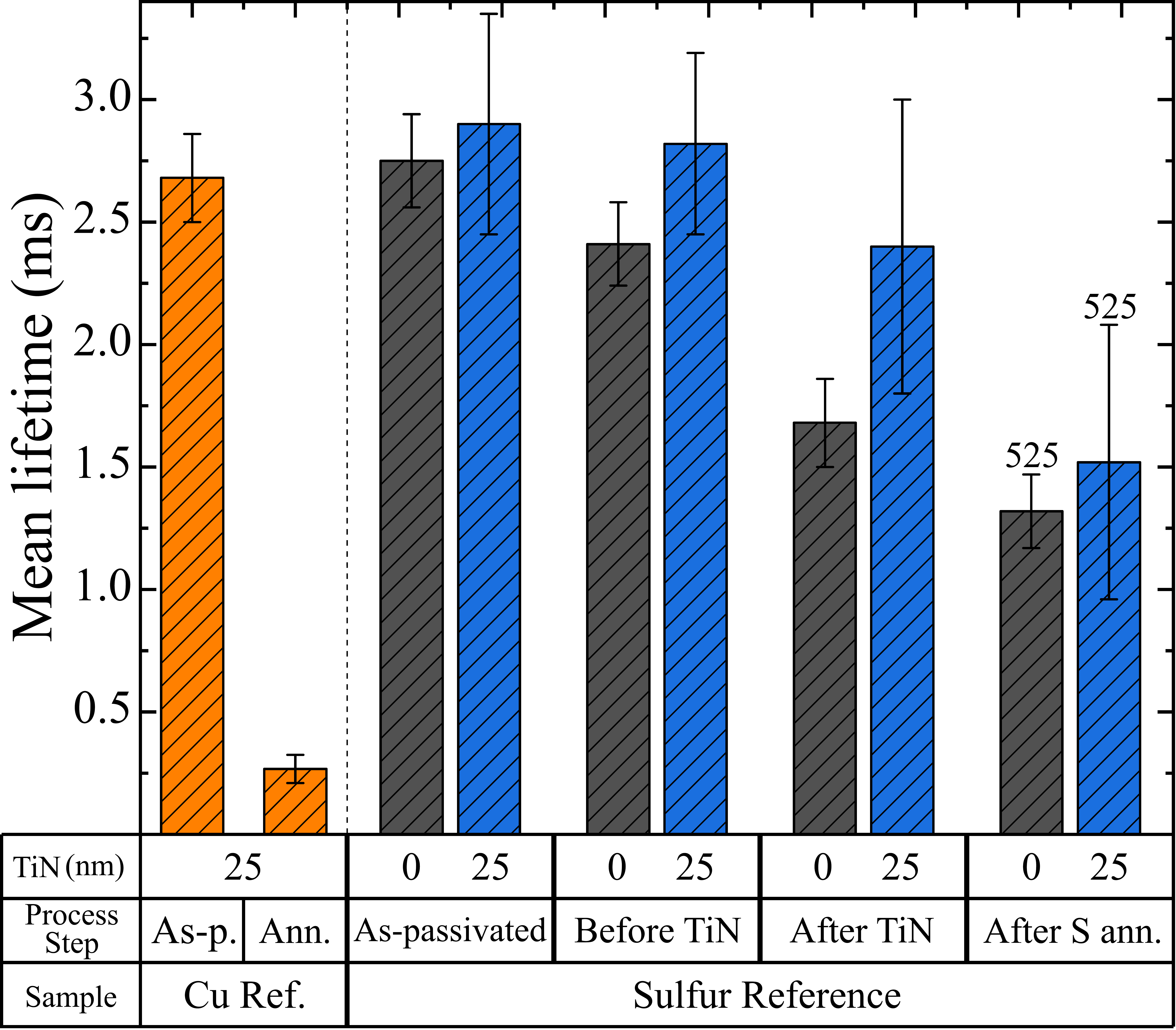}
  \caption{}
  \label{fig:3a}
\end{subfigure} \\
\begin{subfigure}[b]{0.6\textwidth}
  \centering
  \includegraphics[width=\linewidth]{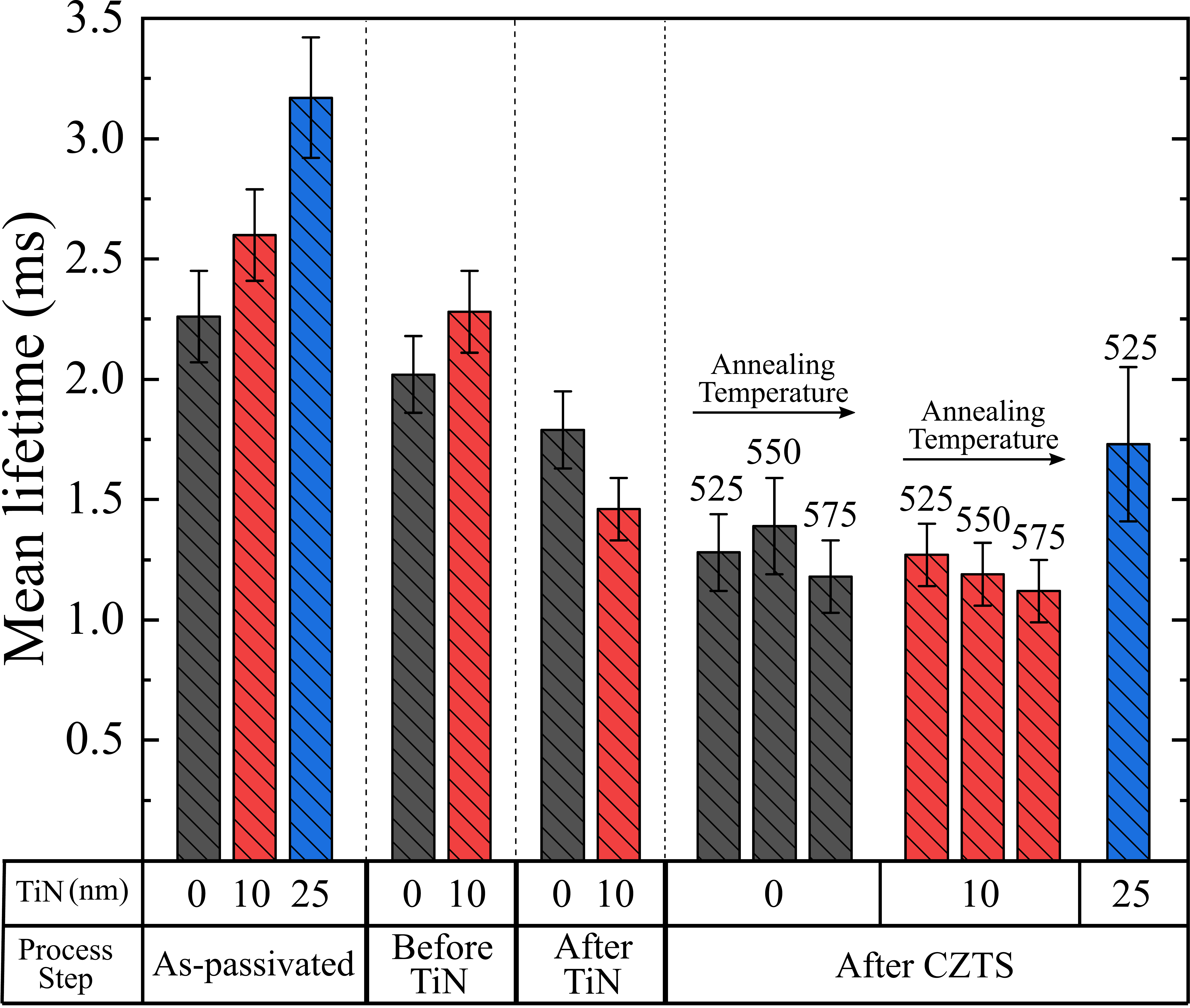}
  \caption{}
  \label{fig:3b}
\end{subfigure}
\caption{Mean effective minority carrier lifetimes of Si for (a) the \textit{Cu Reference} and \textit{Sulfur Reference} samples, and (b) the \textit{CZTS}-processed samples, at each different processing step. The temperatures of the annealing series (in \degree C) are indicated above the corresponding bars. Note: All the samples have 25 nm TiN on the backside, so they all show some ``After TiN"  degradation.}
\label{fig:3}
\end{figure}

In Figure \ref{fig:3} (a) and (b), the Si carrier lifetime results for the \textit{Sulfur Reference} and \textit{CZTS} cases are shown. Here, the lifetime was monitored after each major processing step, namely the TiN deposition and the CZTS annealing steps. From the $``$As-Passivated$"$  to the $``$Before TiN$"$  step, there was a waiting time on the order of a few weeks, causing a slight decrease in the lifetime due to aging. The final lifetimes of Figure \ref{fig:3} (a) are reduced to 45-50\%  of the $``$As-passivated$"$  value after annealing in a sulfur atmosphere, suggesting the role of S as a contaminating species. The 60 min point of the \textit{Cu Reference} carrier lifetime of Figure \ref{fig:2} is included in Figure \ref{fig:3} (a) for comparison, showing that the impact of the S atmosphere is less severe than that of metallic Cu. Further details and lifetime maps of the \textit{Cu Reference} and \textit{Sulfur Reference} carrier lifetime measurements are shown in the supplementary Figure S2 and Figure S3, respectively.

In Figure \ref{fig:3} (b), the key observation is that the final lifetime values after \textit{CZTS} processing are significantly higher than that of the \textit{Cu Reference} case, in spite of the fact that metallic Cu is present in the co-sputtered CZTS precursors. One possible explanation for this milder contamination effect is that the driving force for the formation of Cu\textsubscript{2-x}S phases (the binary phases in the CZTS phase diagram with the lowest melting point \cite{doi:10.1002/9781118437865.ch3}) competes directly with the diffusion of the available Cu into Si. Therefore, the competing Cu\textsubscript{2-x}S formation reaction reduces the driving force for Cu diffusion into Si. Moreover, the lifetime values after \textit{CZTS} processing are comparable to the \textit{Sulfur Reference} case, indicating that having CZTS in addition to a sulfur atmosphere does not lead to additional lifetime deterioration.

\begin{figure}
  \centering
  \includegraphics[width=0.6\linewidth]{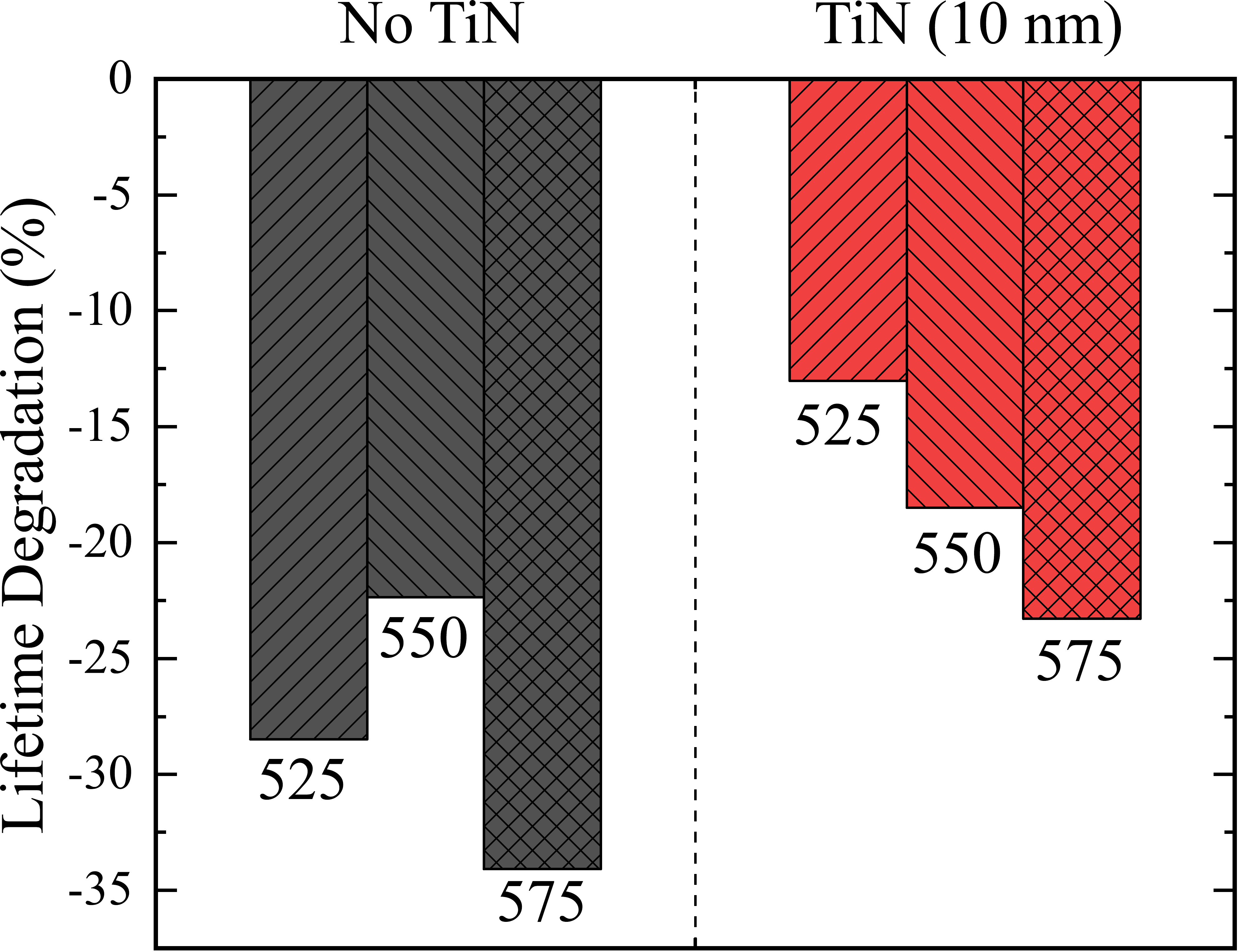}
  \caption{Relative change in minority carrier lifetime for the annealing series in the \textit{CZTS}-processed samples, when scaled to the respective ``After TiN" lifetimes. The temperature is displayed at the bars (in \degree C).}
    \label{fig:4}
\end{figure}

The influence of the annealing temperature was also studied in the \textit{CZTS} case with a series of different annealing temperatures at 525 \degree C, 550 \degree C and 575 \degree C, for the cases without TiN and 10 nm TiN, as shown in the ``After CZTS"  step of Figure \ref{fig:3} (b). One single measurement with a TiN thickness of 25 nm is included as reference for comparison in the subsequent studies. However, this thickness would be too high for use in a tandem cell (due to poor transparency). The annealing series shows that while the 10 nm TiN case seems to follow a trend with increasing temperature, this is not true for the case without TiN. This is likely due to spatial variations in the sample’s lifetime, shown by the uncertainty bars, which have a magnitude comparable to the variations seen in the temperature series. Moreover, it can be seen that the 0 nm and 10 nm TiN series have comparable absolute carrier lifetimes after CZTS processing. To further understand this behavior, we plot in Figure \ref{fig:4} the same results but scaled to the respective ``After TiN"  lifetimes. By doing this, it becomes clear that the lifetime deterioration during CZTS processing is more significant when no TiN is present. This scaling procedure is also justified as it can be noted in both Figure \ref{fig:3} (a) and (b) that the final (post-process) lifetime values are affected by a significant and non-uniform loss in lifetime during the TiN deposition step. The reason for this loss may be attributed to a minor contamination originating from the ALD chamber and stainless steel carrier used during the deposition (e.g. iron contamination). We discuss this issue in greater detail and show additional experimental data in the SI (see Figure S8). However, further future investigation will be required to fully clarify this effect.   

Despite the degradation throughout the processing, the final lifetimes are above 1 ms, which corresponds to an i-\textit{V}\textsubscript{oc} above 700 mV. This encouraging result indicates that the performance of the bottom silicon cell may not necessarily be compromised as a result of the CZTS synthesis. However, given the comparable absolute lifetime values, regardless of the TiN thickness, it is not yet clear from the lifetime results alone whether the use of a TiN barrier layer would be useful.

\begin{figure}
  \centering
  \includegraphics[width=0.6\linewidth]{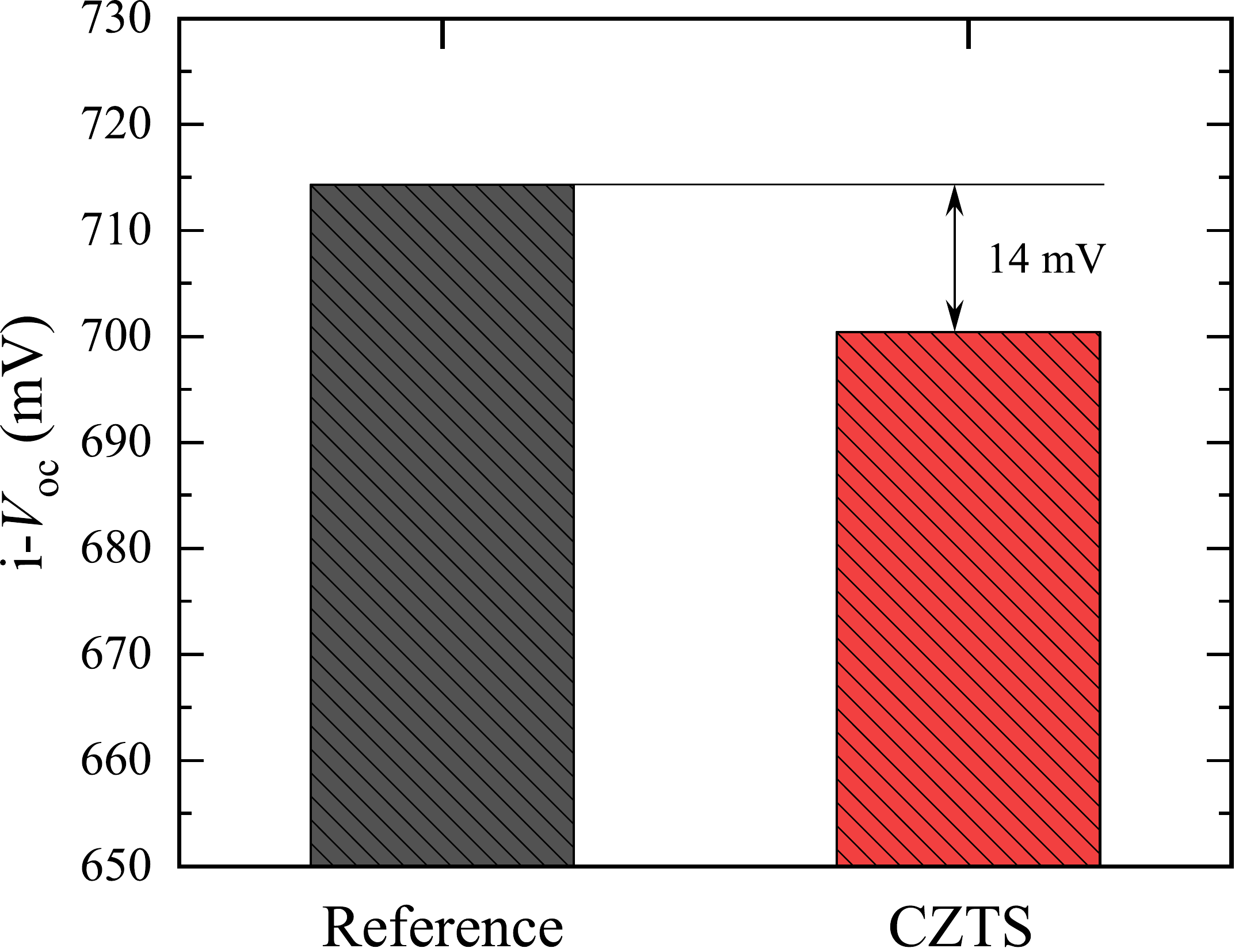}
  \caption{Comparison between the i-\textit{V}\textsubscript{oc} of a \textit{CZTS}-processed sample, annealed in a sulfur atmosphere at 525 \degree C for 30 min (in red), and a reference without CZTS processing (in grey). Both half-wafers are cleaved from the same substrate, and are end-passivated with 40 nm Al\textsubscript{2}O\textsubscript{3} on both sides after etching the TiN and CZTS layers.}
    \label{fig:5}
\end{figure}

To further evaluate whether the observed degradation is related to a bulk contamination or TO/n+PolySi surface depassivation, a complementary experiment was conducted where a silicon wafer was passivated only at the end of the CZTS processing, after etching the CZTS and TiN layers. Any possible unforeseen effects caused by the TO/n+PolySi passivation are avoided by using this configuration. We refer to this as the ``end-passivated" sample. Here, we repeated the Si/TiN(25nm)/CZTS sample, except using a non-passivated bare silicon wafer (no TO/n+PolySi passivation) as the substrate. Subsequent to the CZTS and TiN etching and cleaning, 40 nm ALD Al\textsubscript{2}O\textsubscript{3} was deposited on both sides for surface passivation. The results, plotted in Figure \ref{fig:5}, indicate a tolerable 14 mV decrease in i-\textit{V\textsubscript{oc}} ($\sim$ 30\%  lifetime decrease) for the sample with CZTS processing compared to the clean reference sample.

Even though this experiment does not directly clarify the effect of the PolySi layer on the diffusion of contaminants from CZTS processing, it shows that relatively high-end lifetimes can be achieved without using a PolySi layer. This suggests that there is some flexibility of design in the bottom Si cell, and offers new perspectives for future tandem integration experiments.

\subsection{SIMS and RBS Analysis}

\begin{figure}
\centering
\begin{subfigure}[b]{.31\textwidth}
  \centering
  \includegraphics[width=\linewidth]{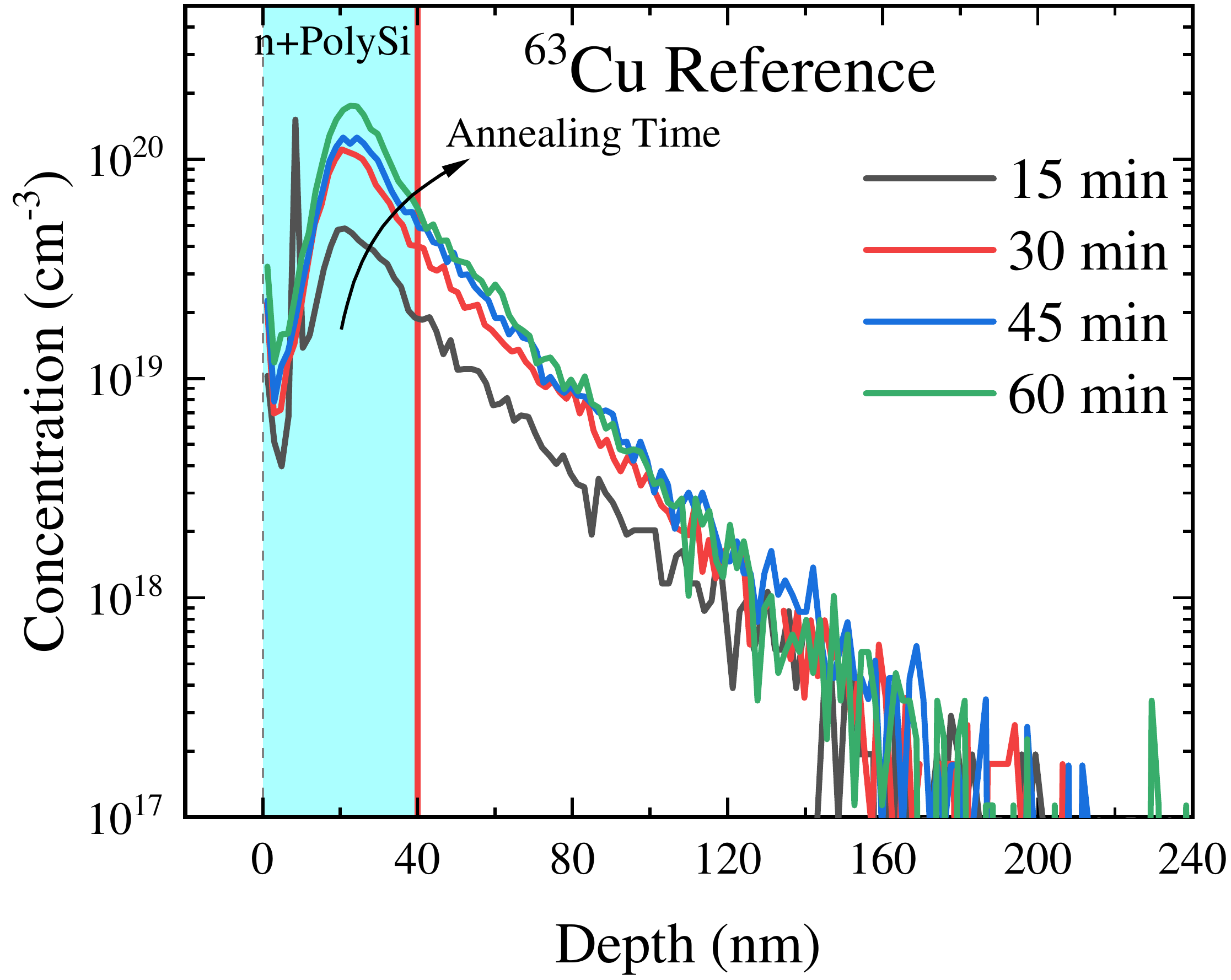}
  \caption{}
  \label{fig:6a}
\end{subfigure}
\begin{subfigure}[b]{.31\textwidth}
  \centering
  \includegraphics[width=\linewidth]{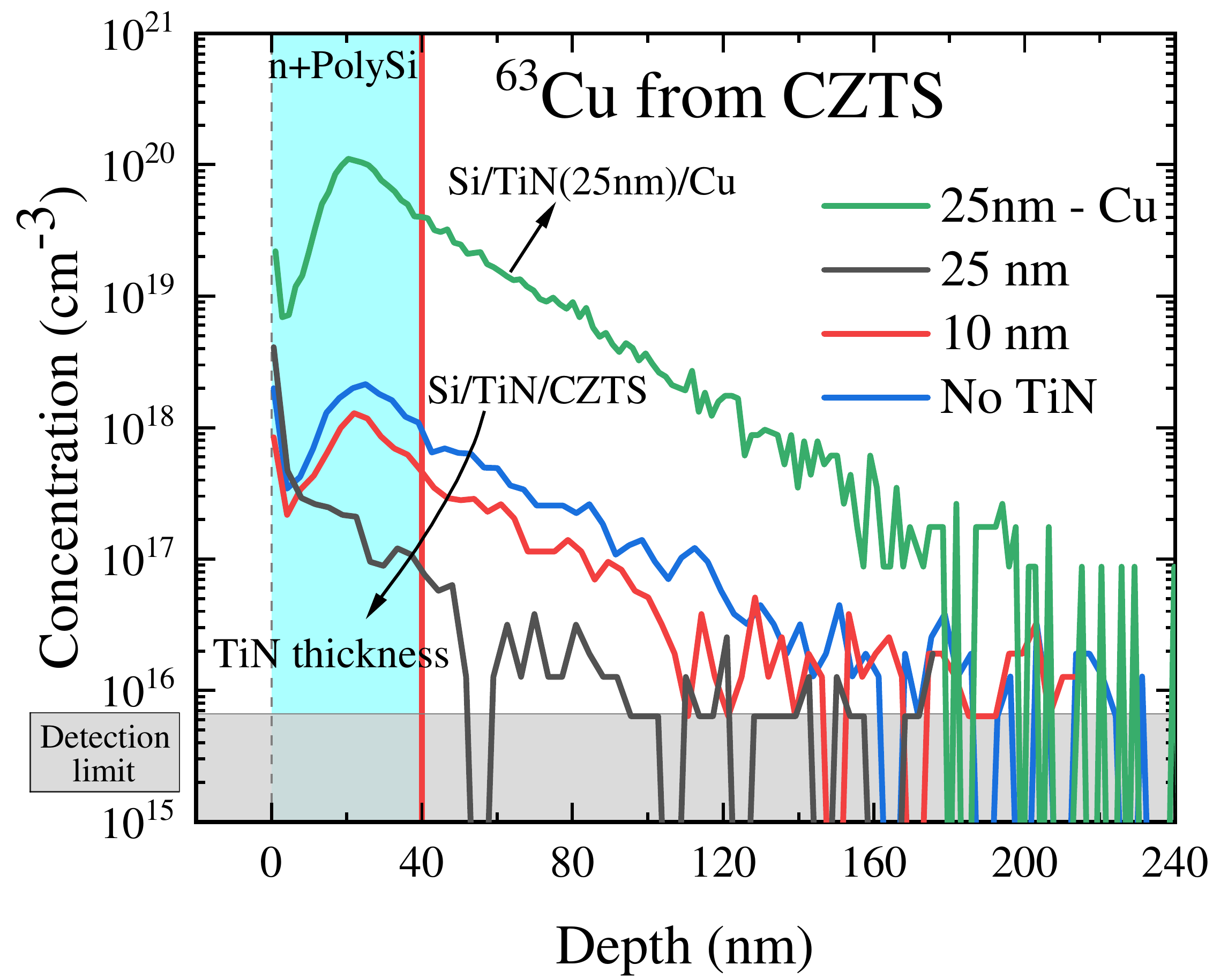}
  \caption{}
  \label{fig:6b}
\end{subfigure}
\begin{subfigure}[b]{.31\textwidth}
  \centering
  \includegraphics[width=\linewidth]{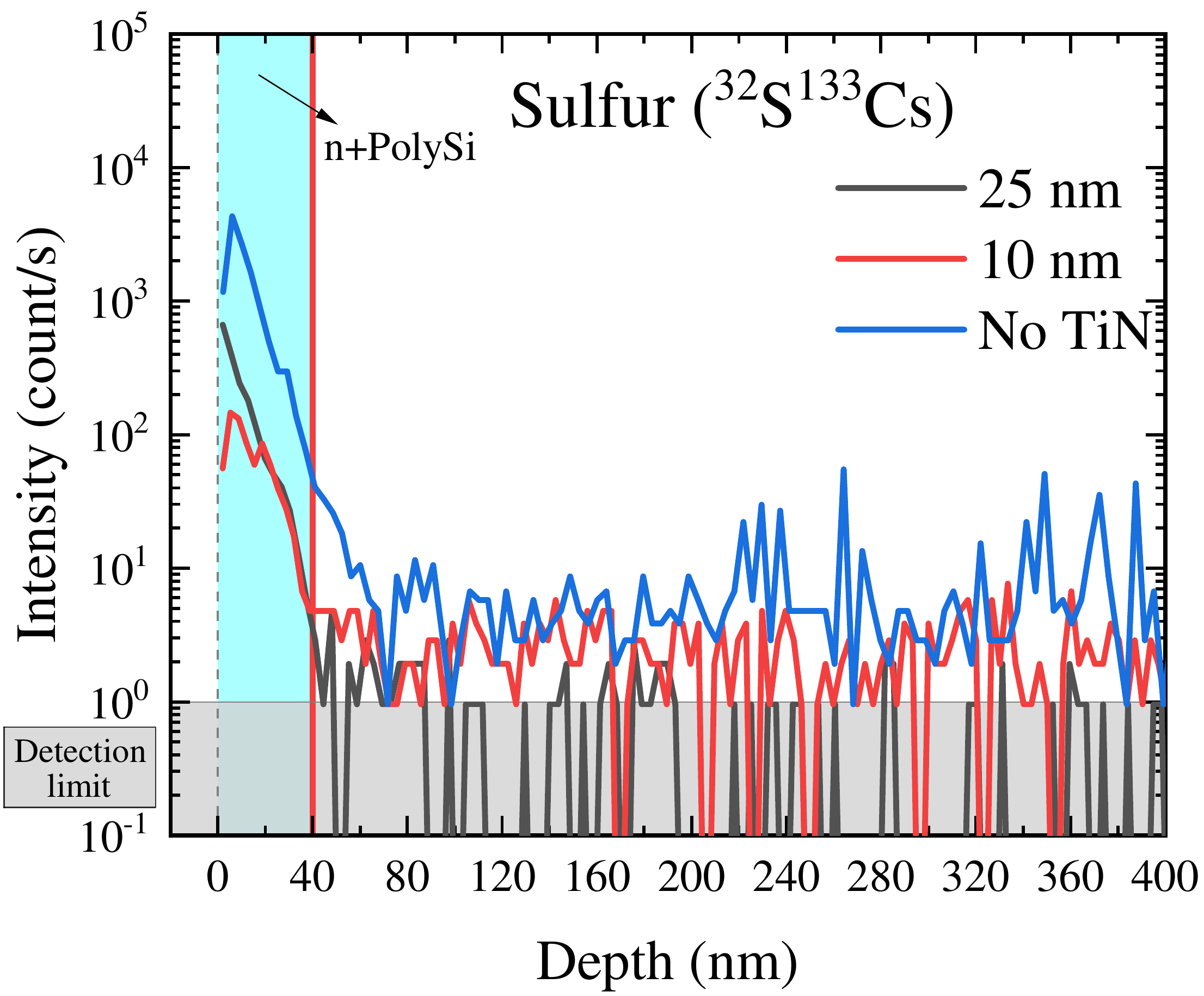}
  \caption{}
  \label{fig:6c}
\end{subfigure}\\
\begin{subfigure}[b]{.31\textwidth}
  \centering
  \includegraphics[width=\linewidth]{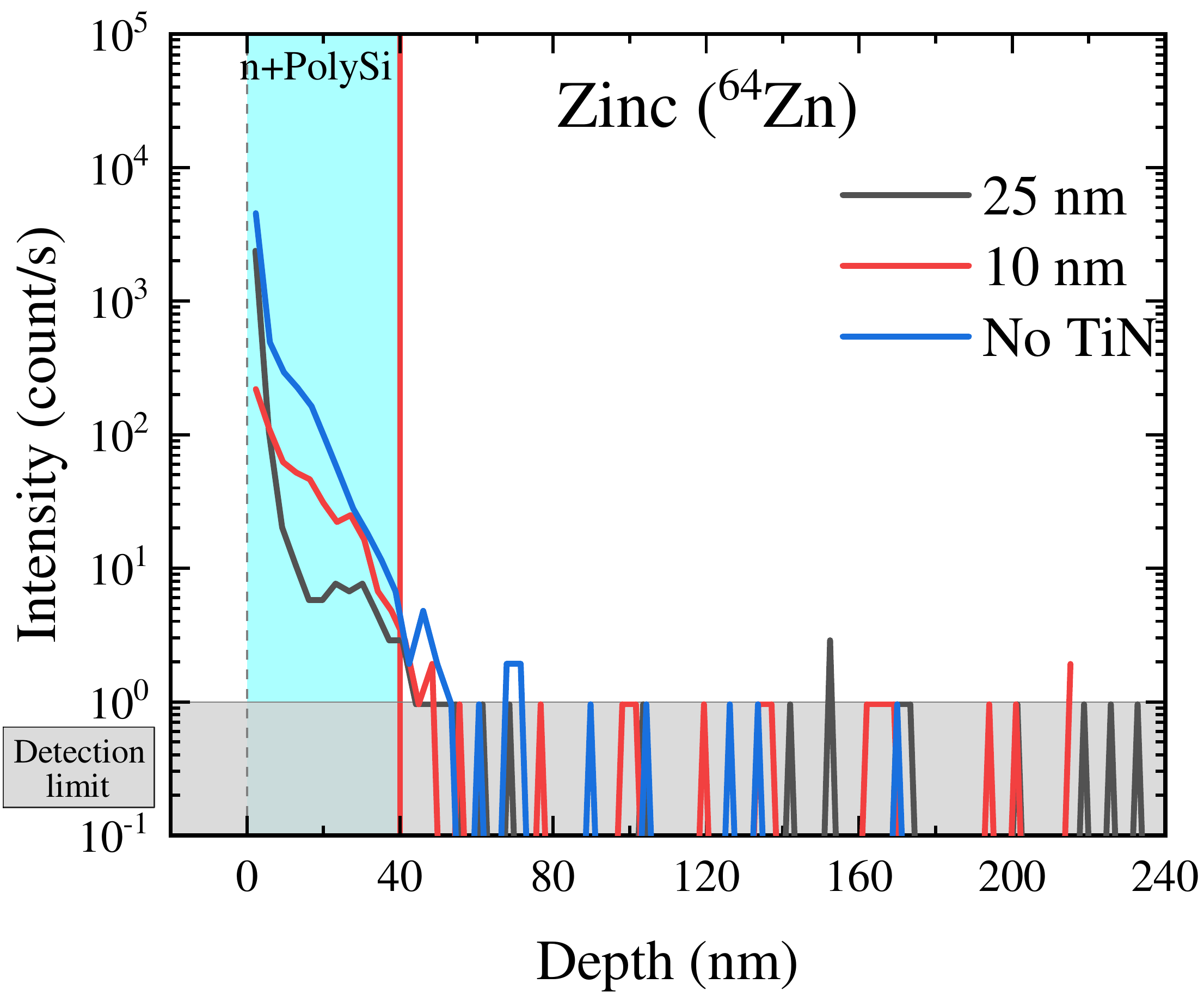}
  \caption{}
  \label{fig:6d}
\end{subfigure}
\begin{subfigure}[b]{.31\textwidth}
  \centering
  \includegraphics[width=\linewidth]{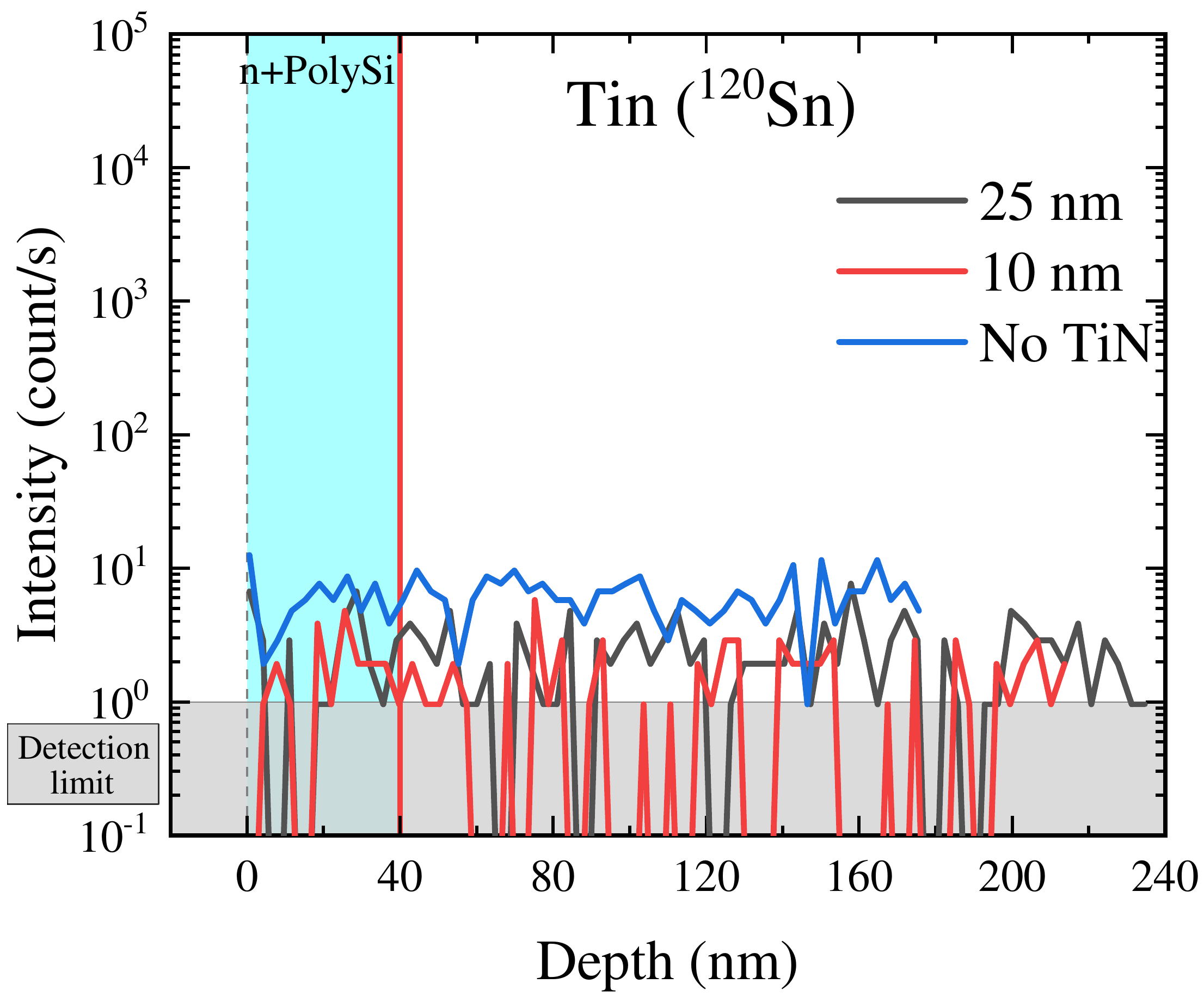}
  \caption{}
  \label{fig:6e}
\end{subfigure}
\begin{subfigure}[b]{.31\textwidth}
  \centering
  \includegraphics[width=\linewidth]{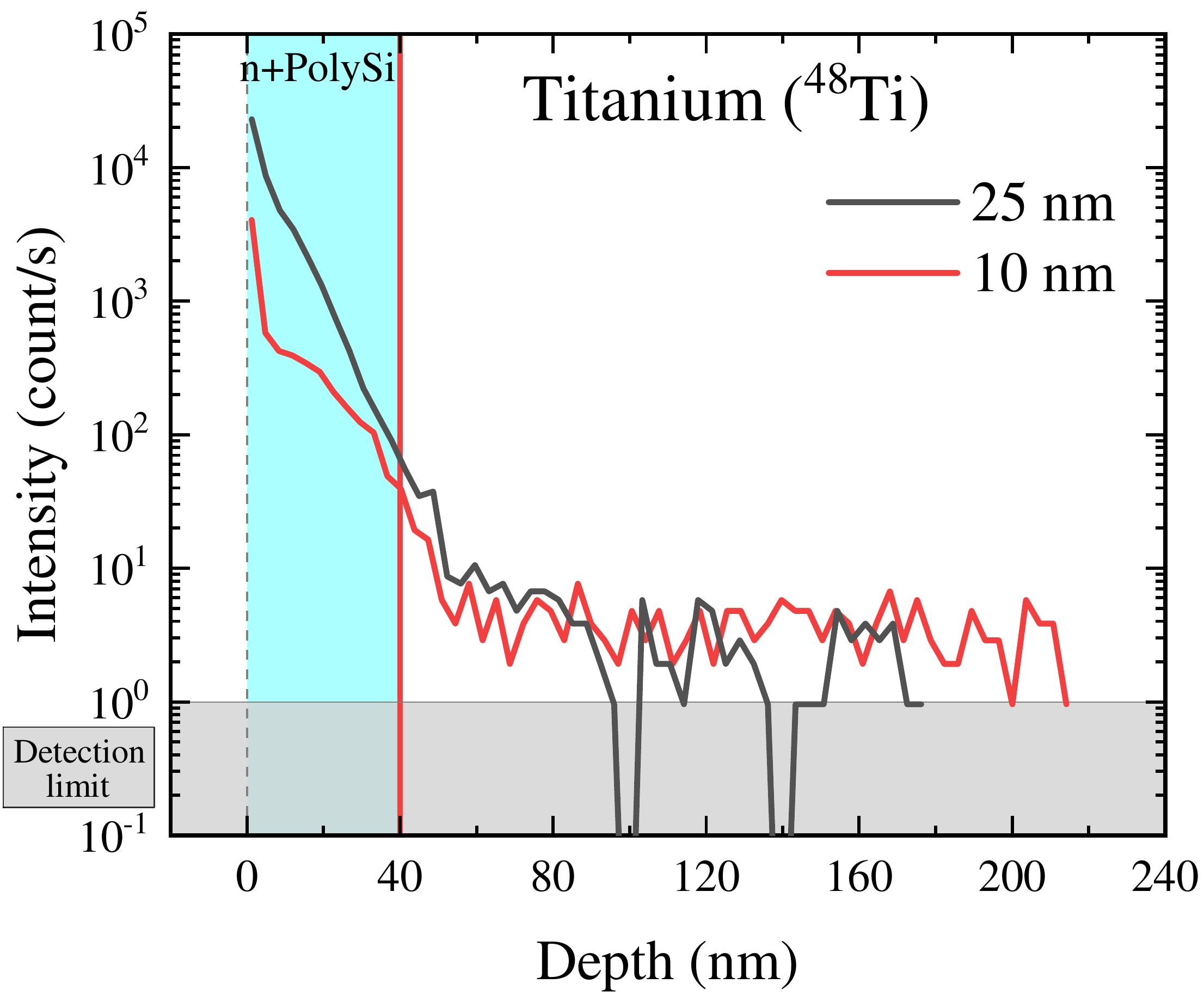}
  \caption{}
  \label{fig:6f}
\end{subfigure}
\caption{SIMS depth profiles for \textit{Cu Reference} and \textit{CZTS}-processed samples. (a) The \textit{Cu Reference}, showing quantitative Cu depth profiles; (b) Quantitative Cu depth profile for the \textit{CZTS}-processed samples. The depth profile of the \textit{Cu Reference} sample is added for comparison; (c), (d), (e) and (f) Qualitative depth profiles of Zn, Sn, S, and Ti for the CZTS samples, respectively. The measurements are performed on the n+PolySi layer towards the c-Si bulk, as marked by the blue rectangle, after etching the top layers. The \textit{CZTS}-processed samples were annealed at 525 \degree C. The annealing time was 30 min unless otherwise specified.}
\label{fig:6}
\end{figure}

To correlate the lifetime results with possible diffusion of contaminants into the Si bulk, SIMS and RBS measurements were performed on selected \textit{Cu Reference} and \textit{CZTS}-processed samples (after selective removal of the Cu, TiN and CZTS layers). The SIMS results are illustrated in Figure \ref{fig:6}. For the \textit{Cu Reference} samples, the corresponding quantitative Cu SIMS depth profiles are shown in Figure \ref{fig:6} (a). A clear diffusion tail into the c-Si bulk is detected in all cases, with a Cu peak concentration of up to 10\textsuperscript{20} cm\textsuperscript{-3} occurring in the PolySi. Furthermore, an increase in Cu concentration is seen with increasing annealing time, which is in qualitative agreement with the lifetime results of Figure \ref{fig:2}. In Figure \ref{fig:6} (b), a quantitative Cu profile is presented for the \textit{CZTS}-processed samples annealed at 525 \degree C. The Cu profiles reveal that for the No TiN and 10 nm TiN samples, there is a diffusion tail extending at least 100 nm into the Si bulk, but for the 25 nm TiN case the Cu concentration drops sharply to below detection limits after the PolySi. In all three cases, the Cu concentration is 2 -- 3 orders of magnitude lower compared to the \textit{Cu Reference} case, which helps to justify their significantly higher lifetimes. Into the Si bulk (close to the surface), the Cu concentration is always lower than 10\textsuperscript{18} cm\textsuperscript{-3}, which in Si corresponds to 0.002 at\% (or 20 ppm).

Depth profiles of other relevant elements during CZTS processing, namely Zn, Sn, S and Ti (from TiN) are shown in Figure \ref{fig:6} (c) to (f). Other elements than Cu appear to be at background levels or near detection limits into the c-Si bulk.

\begin{figure}
\centering
\begin{subfigure}[b]{0.6\textwidth}
  \centering
  \includegraphics[width=\linewidth]{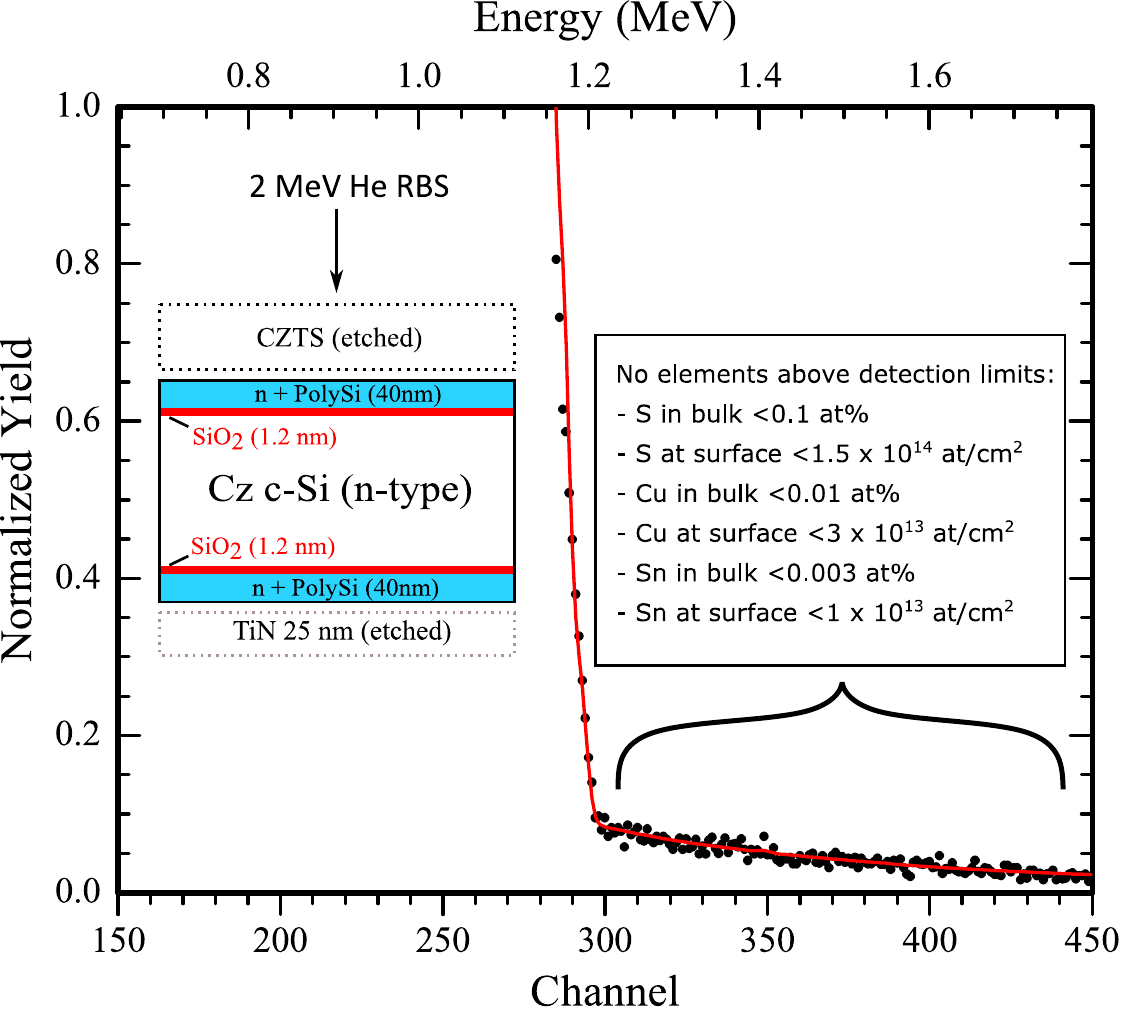}
  \caption{}
  \label{fig:7a}
\end{subfigure} \\
\begin{subfigure}[b]{.6\textwidth}
  \centering
  \includegraphics[width=\linewidth]{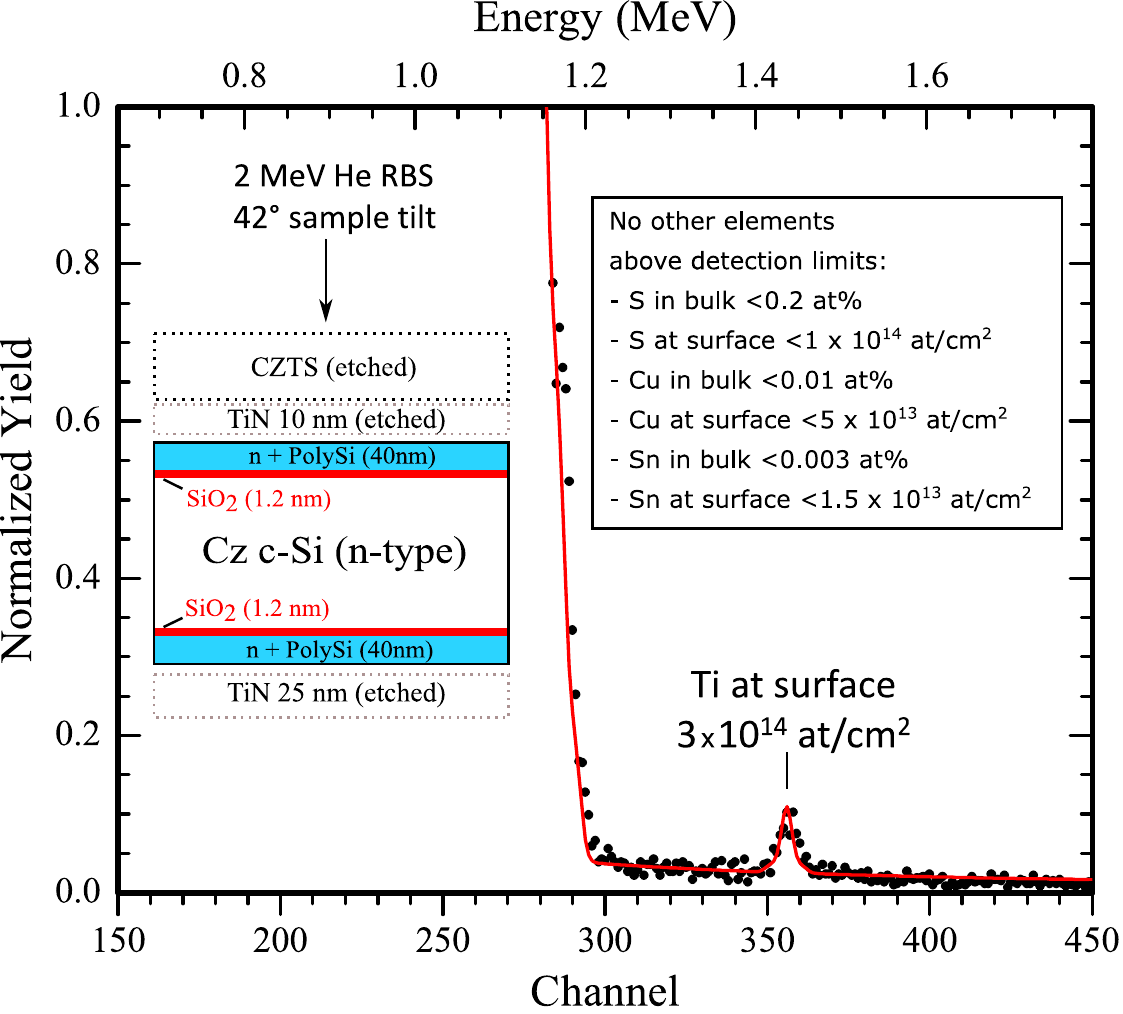}
  \caption{}
  \label{fig:7b}
\end{subfigure}
\caption{RBS spectra for the \textit{CZTS}-processed samples with (a) No TiN and (b) 10 nm TiN. The insets show the structures used and the detection limits for the contaminant elements. The samples were annealed at 525 \degree C.}
\label{fig:7}
\end{figure}

To complement the SIMS analysis, RBS measurements were done on the \textit{CZTS}-processed samples with 0 nm and 10 nm TiN, annealed at 525 \degree C. The results are illustrated in Figure \ref{fig:7} (a) and (b), respectively. Since all the potential contaminant elements are heavier than Si, the RBS data is zoomed in at energies higher than the Si onset. None of the possible contaminants are detected in Si, except for some Ti at the surface, which was not fully removed during the piranha etching (confirmed with SEM, not shown here). This means that an estimate for the upper limit for the concentration of these contaminants can be established, given by the sensitivity of the measurement itself. For the measurement conditions of Figure \ref{fig:7} (a) and (b), these upper limits are given in the figure insets. In the particular case of Cu, which was also quantitatively measured by SIMS, RBS shows that its concentration has to be below 0.01 at\%, agreeing with the value of below 0.002 at\%  obtained by SIMS.

\subsection{DLTS Analysis}

\begin{figure}
  \centering
  \includegraphics[width=0.6\linewidth]{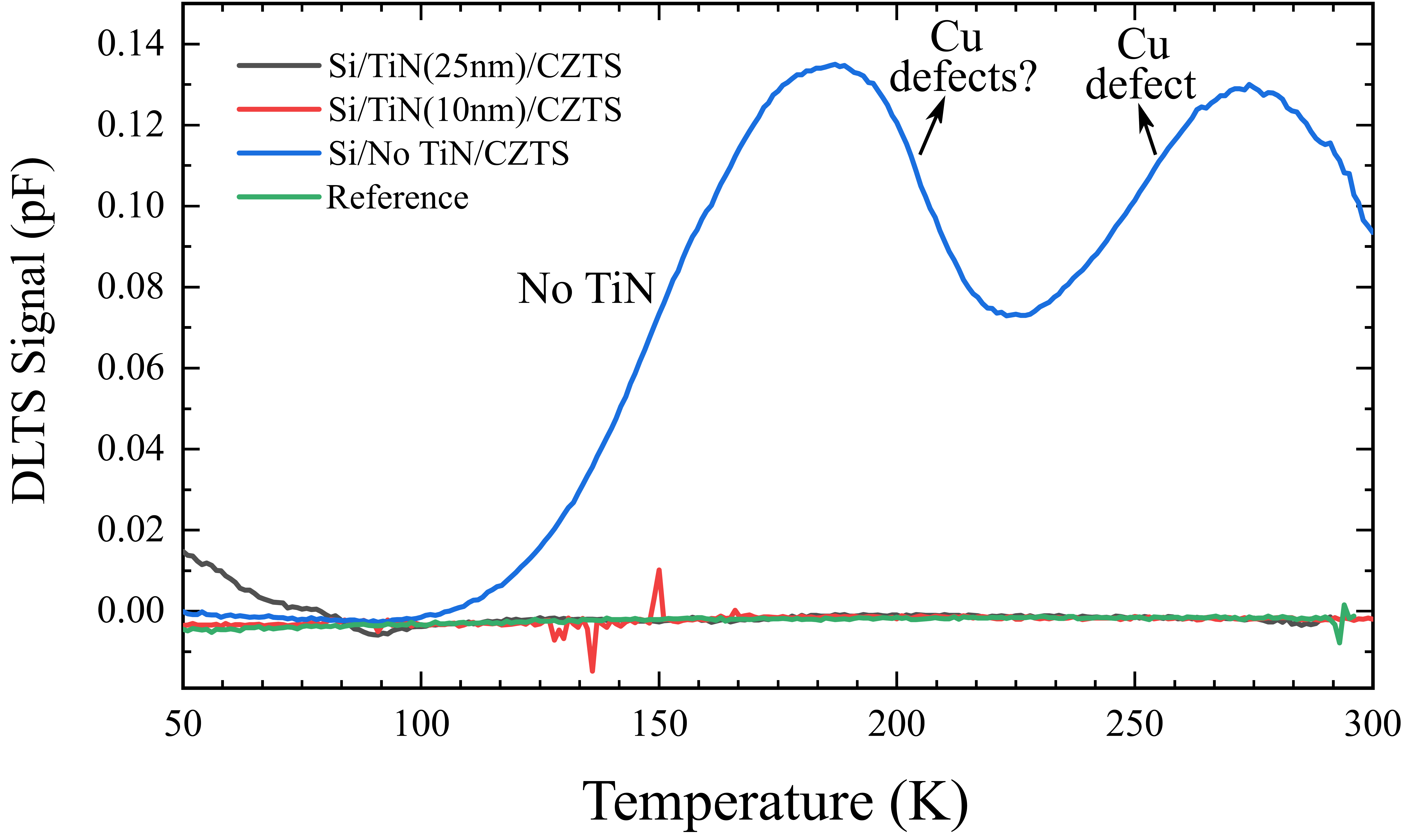}
  \caption{DLTS results of \textit{CZTS}-processed samples annealed at 525 \degree C compared to a clean reference Si wafer. Note: these DLTS samples do not have the TO/n+PolySi passivation stack.}
    \label{fig:8}
\end{figure}

To further assess the influence of these possible contaminants, DLTS measurements were made on \textit{CZTS}-processed samples annealed at 525 \degree C. Here, samples without the TO/n+PolySi passivation were prepared as no Schottky contact (required in our DLTS setup) could be obtained between the metal electrode and the heavily-doped PolySi. An unprocessed bare ``Reference"  wafer was also included to rule out any possible pre-existing defects. The results are plotted in Figure \ref{fig:8}. It is shown that the samples with 10 nm TiN, 25 nm TiN and the Reference wafer do not have any DLTS signal, but the No TiN sample exhibits peaks related to electrically active defects, with two features peaking at  $\sim$ 175 K and $\sim$ 275 K. The 175 K peak shows a broadening towards the lower temperature side, which may be related to several overlapping defect signatures or extended defects \cite{doi:10.1002/(SICI)1099-159X(199703/04)5:2<79::AID-PIP155>3.0.CO;2-J}. In the case of extended defects, an exponential decay in emission rate may not hold, and will influence the extracted activation energies and apparent capture cross-sections from an Arrhenius plot of the corresponding DLTS peak \cite{DOOLITTLE1986227}. This peak near 175 K might come from several defects associated with precipitates of Cu \cite{Haase_2017, doi:10.1063/1.344389}, but further measurements would be required to assign this unambiguosly. The peak at 275 K could be used instead for making an Arrhenius plot. This peak has a broad shape due to its very low capture cross-section of $2\times10^{-22}$ cm\textsuperscript{2}, and its energy level was found to be 0.16 eV below the conduction band edge, as extracted from the Arrhenius plot. The level at  $E_{c}-0.16$ eV has previously been reported in Cu diffused Si and shown to originate from interstitial copper or a complex of interstitial copper by Istratov et al \cite{PhysRevB.52.13726}. More details on the DLTS results, analysis and Arrhenius plot can be found in the supplementary information (Figure S4 and Figure S5).

Based on these findings, 10 nm of TiN seems to be sufficient to prevent the formation of electrically active defects in the Si bulk. This thickness was thus selected to prepare a full CZTS/Si tandem solar cell.

\subsection{Fabrication of a Monolithic CZTS/Si Solar Cell}\label{sec:34}

\begin{figure}
  \centering
  \includegraphics[width=0.6\linewidth]{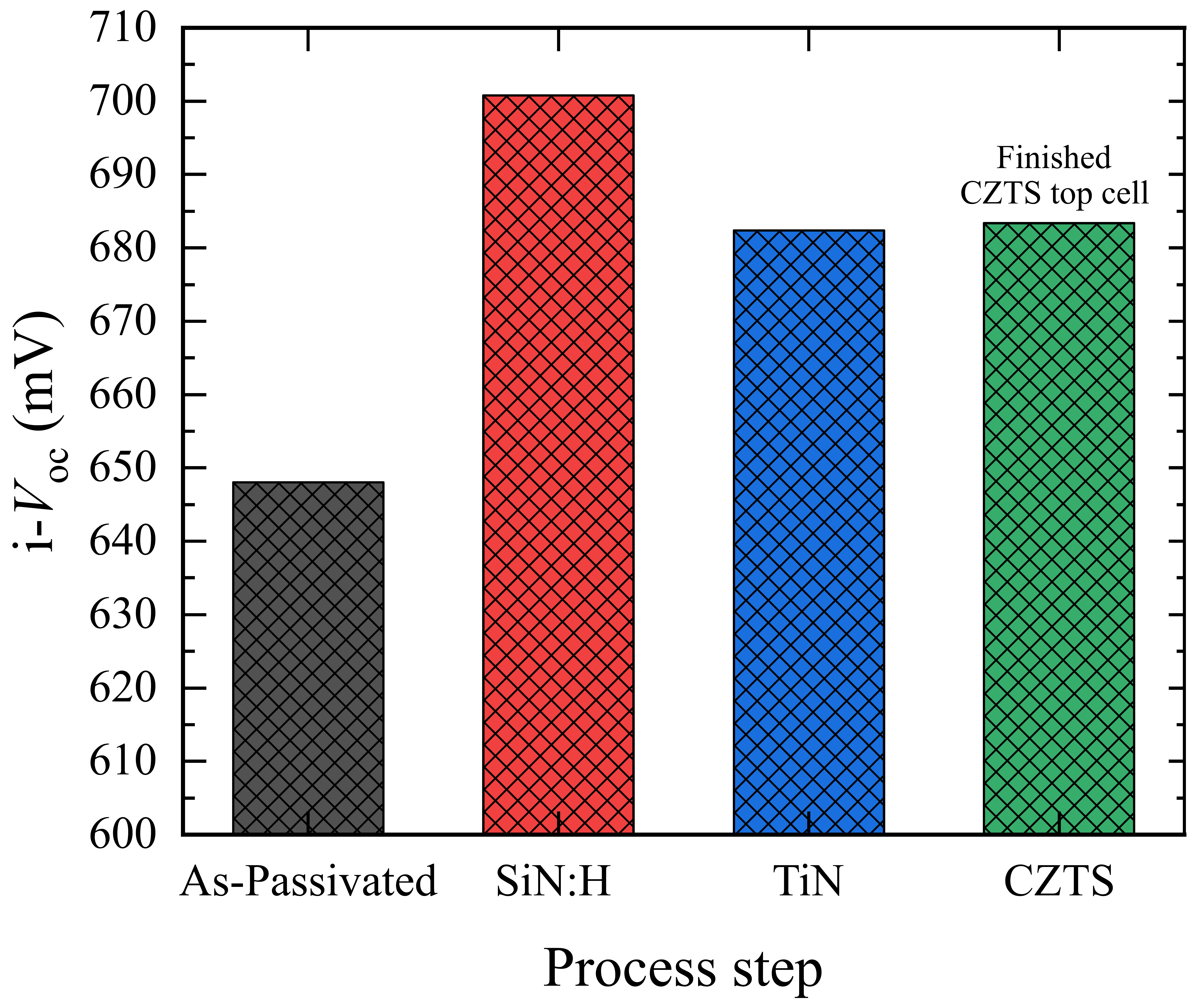}
  \caption{Changes in i-\textit{V}\textsubscript{oc} of the silicon bottom cell during the fabrication processes of the CZTS/Si tandem cell. The indicated fabrication steps are: 1) Silicon surface passivation 2) SiN:H hydrogenation of the TOPCon layers 3) TiN deposition, and 4) Full CZTS cell fabrication before depositing the Ag back contact.}
    \label{fig:9}
\end{figure}

\begin{figure}
\centering
\begin{subfigure}{.36\textwidth}
  \centering
  \includegraphics[width=\linewidth]{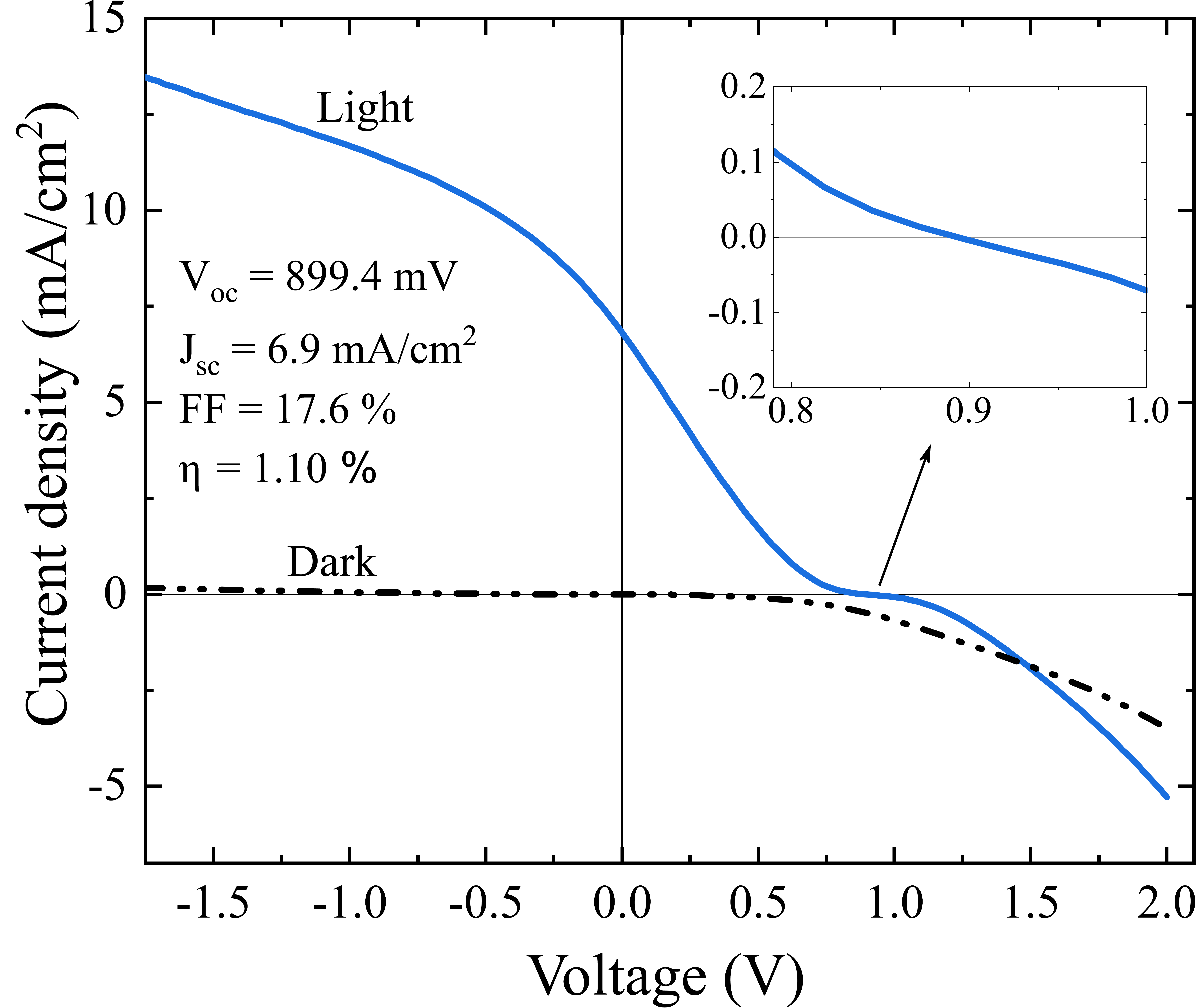}
  \caption{}
  \label{fig:10a}
\end{subfigure}
\begin{subfigure}{.37\textwidth}
  \centering
  \includegraphics[width=\linewidth]{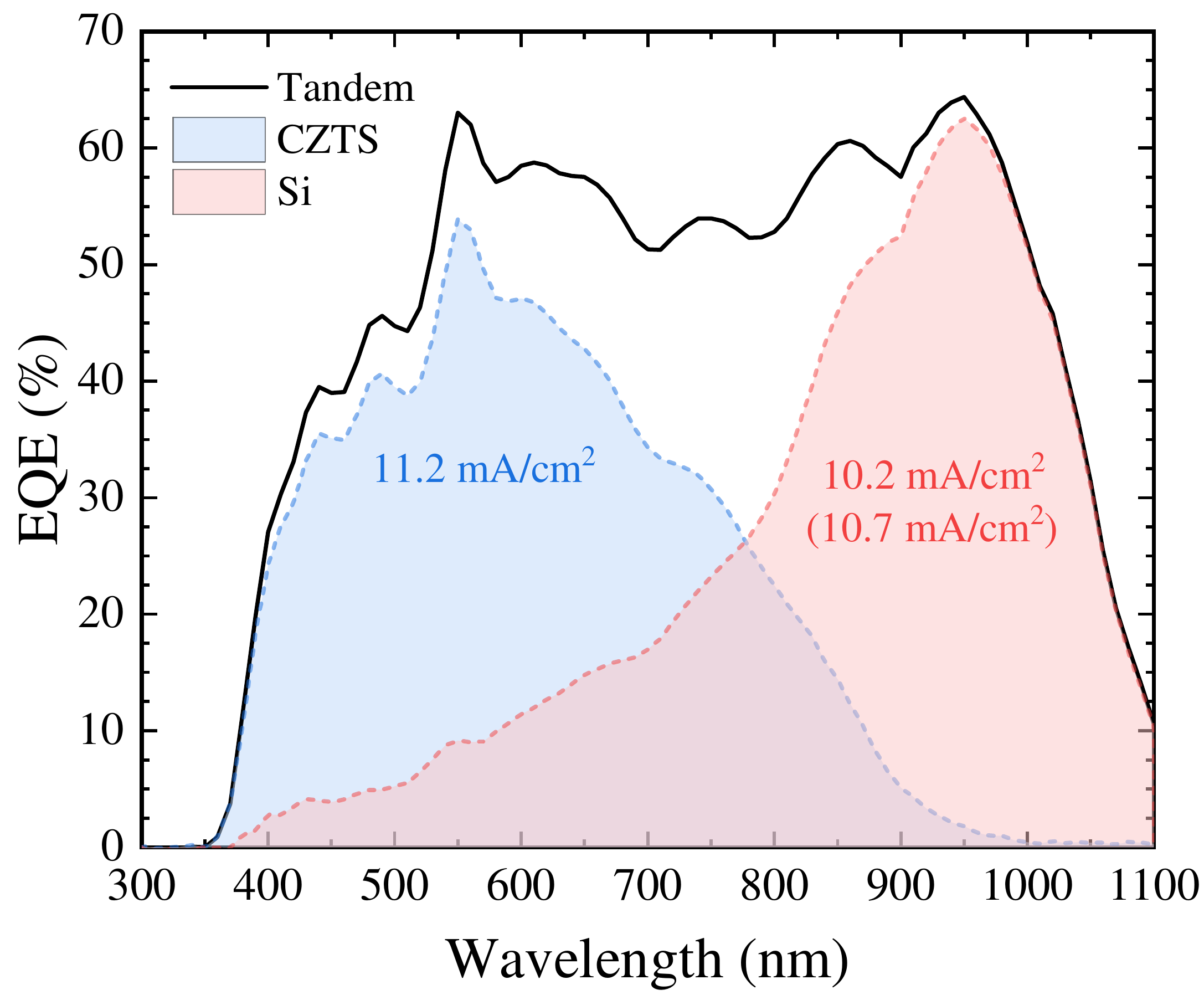}
  \caption{}
  \label{fig:10b}
\end{subfigure}
\begin{subfigure}{.2\textwidth}
  \centering
  \includegraphics[width=\linewidth]{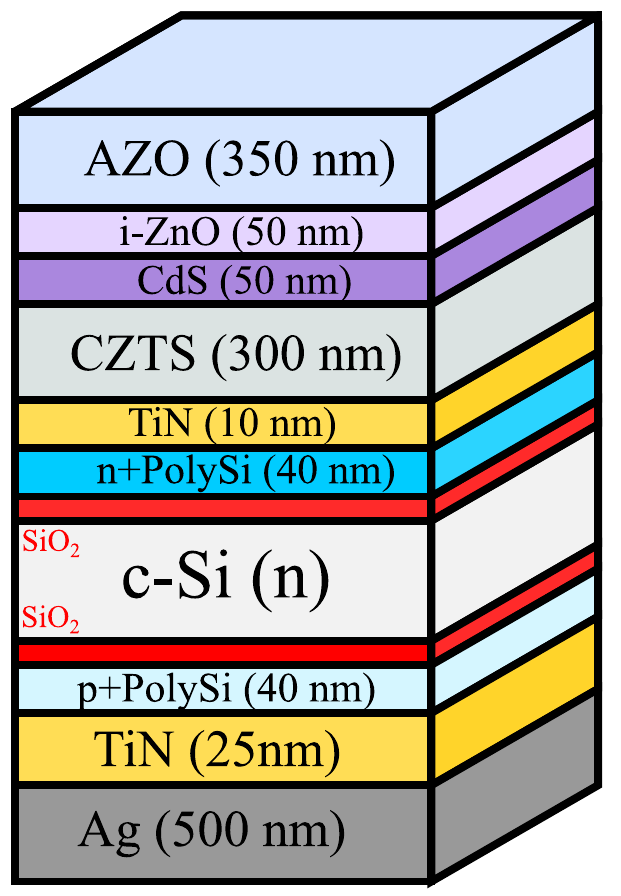}
  \caption{}
  \label{fig:10c}
\end{subfigure}
\caption{(a) Tandem cell illuminated (solid) and dark (dashed) J-V curves. The insets show the region near V\textsubscript{oc} and the J-V parameters; (b) EQE of the two sub-cells and sum of the sub-cell contributions. The value in parenthesis accounts for a $\sim$ 0.5 mA/cm\textsuperscript{2} contribution from the high wavelength region outside the measurement range; (c) Tandem solar cell scheme}
\label{fig:10}
\end{figure}

The effective minority carrier lifetime of the silicon bottom cell was monitored at different steps of the fabrication process, similar to Section \ref{sec:31}. However, the samples are now asymmetrically passivated (with TO/p+PolySi on the backside). An additional SiN hydrogenation step (to improve the passivation quality) was also included. The corresponding i-\textit{V}\textsubscript{oc} is shown in Figure \ref{fig:9} as a function of process steps. Figure \ref{fig:9} shows that the i-\textit{V}\textsubscript{oc} of the silicon bottom cell was slightly degraded after the TiN deposition step. As mentioned in Section \ref{sec:31}, we suggest that this degradation may be due to iron contamination (originating from the ALD chamber). However, given the additional hydrogenation step of the p-PolySi explained above, a partial loss in the hydrogen passivation could occur in this case (see supplementary Figure S8). The i-\textit{V}\textsubscript{oc}, however, does not degrade further during the full fabrication of the CZTS top cell. This demonstrates that 10 nm TiN was an effective diffusion barrier. The J-V curves, EQE and schematic illustration of the tandem device are shown in Figure \ref{fig:10} (a), (b) and (c), respectively.

As seen in Figure \ref{fig:10} (a), the tandem cell shows a \textit{V}\textsubscript{oc} of 900 mV, which is higher than that expected for each individual junction separately, under the same conditions (see supplementary Figure S6). The efficiency, however, is low (1.10\%) and the light J-V curve shows a clear ``rollover" effect, which is characterized by a distortion of the J-V curve, causing a very low fill-factor (FF). The magnitude of the \textit{J}\textsubscript{sc} seems to be also affected by the distortion as the EQE measurement in Figure \ref{fig:10} (b) shows a \textit{J}\textsubscript{sc} of around 11 mA/cm\textsuperscript{2} for each individual cell. We attribute the low efficiency to a combination of the rollover effect and a poorer CZTS top cell compared to the single junction CZTS cell, as will be elaborated below.

This rollover effect, with S-shaped J-V curves, has been reported previously for non-optimal tandem cells, associated with the reverse breakdown voltage regime of the top cell when the tandem is current-mismatched \cite{Torazawa01012016}. While this explanation is certainly plausible here, we note that other effects may cause a rollover effect in single-junction solar cells, as was reviewed recently in \cite{Saive2019}. This occurs when there is one or more barriers to current extraction throughout the solar cell under illumination. This barrier can be due to the presence of Schottky barriers in non-ohmic contacts (namely the n+PolySi/TiN or TiN/CZTS interfaces), or a non-ideal p-n junction, leading to a voltage-dependent current blocking behavior. In particular, this effect has been reported for non-ideal p-n junctions in single-junction CZTS cells \cite{Nakamura_2010}, and we have also seen it in our internal experiments for CZTS/CdS p-n junctions when the synthesis parameters were not ideal (see supplementary Figure S9). For the tandem cell, our TiN barrier layer was produced in an unpassivated ALD chamber, which improves the optical transparency of TiN by incorporating some oxygen, leading to a TiO\textsubscript{x}N\textsubscript{y} film. However, an excessive amount of oxygen can be detrimental, as it is known to increase the sheet resistance of TiN \cite{Maenpaa1980}. It has been reported that the presence of 10-15 \% oxygen in TiN leads to formation of a Schottky diode with a barrier height of 0.55 eV on n-type Si (100) \cite{Ang1988}. In the case of a single Si cell, we found evidence that using a similar 10 nm TiO\textsubscript{x}N\textsubscript{y} layer in-between the n+PolySi and TCO layers can cause a roll-over behavior on single junction Si cells (see supplementary Figure S10). Thus, the results of this work suggest that the ideal compromise between transparency and electrical properties in the TiN layer might not have been reached in this initial device, and this will be investigated in future work by tuning the TiO\textsubscript{x}N\textsubscript{y} composition. Furthermore, a rollover effect has been reported in CIGS at the Mo/CIGS interface on non-glass substrates, where there is no natural inclusion of Na (or other alkali elements) in the absorber layer. It was shown that this effect can be completely eliminated by providing a sufficient amount of Na \cite{doi:10.1063/1.120026}. This is particularly relevant in this work, as the growth of CZTS is also substrate dependent, and no intentional Na was added in the fabrication of the tandem cell.

\begin{figure}
\centering
\begin{subfigure}[b]{.48\textwidth}
  \centering
  \includegraphics[width=\linewidth]{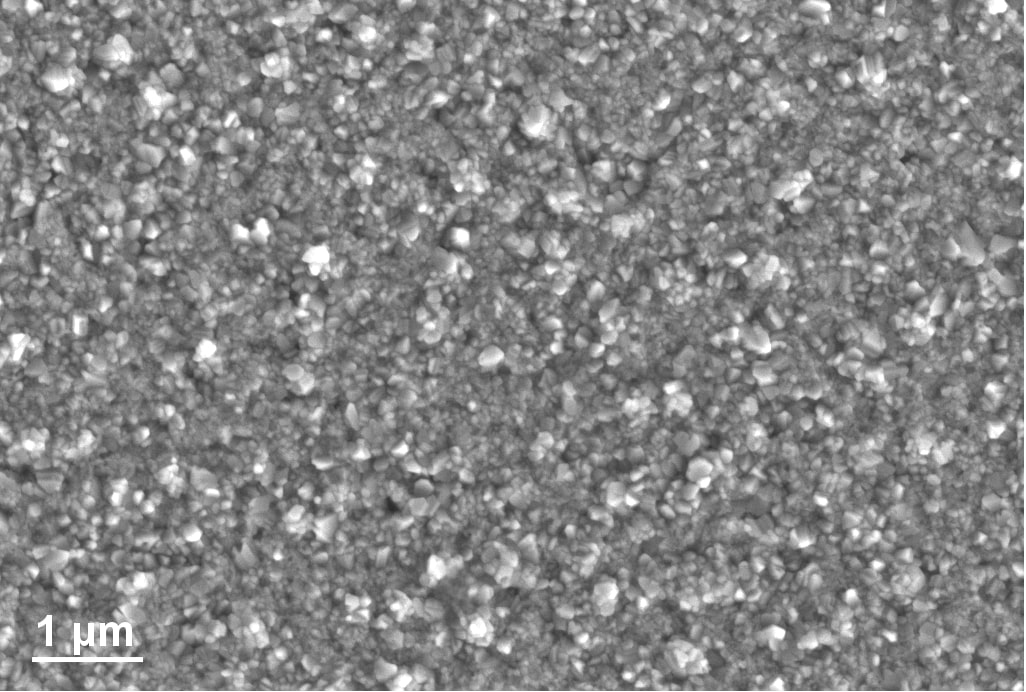}
  \caption{}
  \label{fig:11a}
\end{subfigure}
\begin{subfigure}[b]{.48\textwidth}
  \centering
  \includegraphics[width=\linewidth]{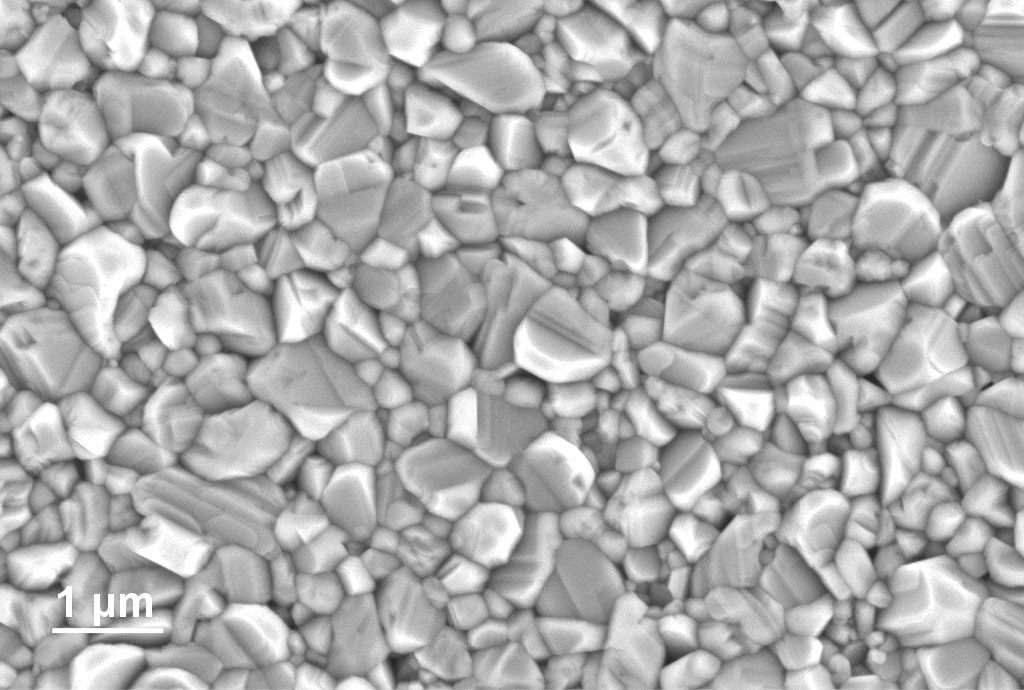}
  \caption{}
  \label{fig:11b}
\end{subfigure}\\
\begin{subfigure}[b]{.48\textwidth}
  \centering
  \includegraphics[width=\linewidth]{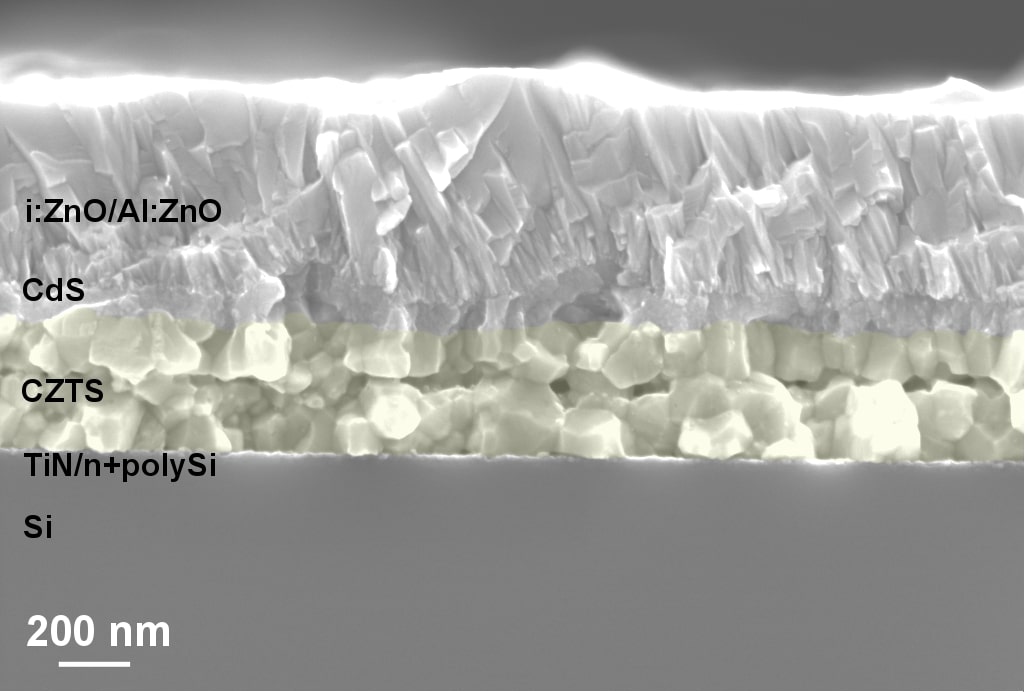}
  \caption{}
  \label{fig:11c}
\end{subfigure}
\begin{subfigure}[b]{.48\textwidth}
  \centering
  \includegraphics[width=\linewidth]{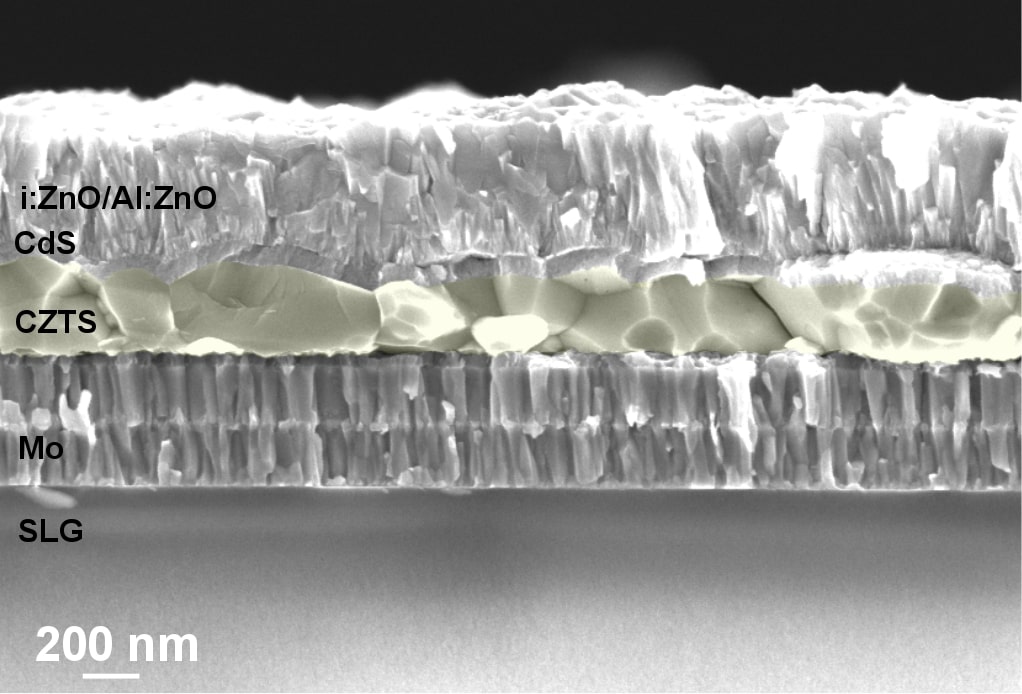}
  \caption{}
  \label{fig:11d}
\end{subfigure}
\caption{SEM comparison of the tandem cell and a single junction thin CZTS cell on Mo/SLG. (a) Top view of the CZTS surface as used on the tandem cell, before CdS deposition; (b) Top view of the thin CZTS surface, before CdS deposition; (c) Cross-section view of the upper part of the tandem; (d) Cross-section view of the full thin CZTS cell. The CZTS absorber layer is highlighted in yellow.}
\label{fig:11}
\end{figure}

To explore these issues, we compare the results of our tandem device to a baseline single-junction CZTS cell where the CZTS thickness was reduced from the typical 1 $\mu$m to a value of around 275 nm, similar to the value used for the tandem. This ``thin CZTS cell" achieved an efficiency of 5.8\%, with a \textit{J}\textsubscript{sc} of 15.8 mA/cm\textsuperscript{2} and a \textit{V}\textsubscript{oc} of 585 mV (the J-V curve is shown in the supplementary Figure S11), which is fairly comparable to state of the art thin CZTS devices (with a record of 8.57\%  for a 400 nm thick CZTS \cite{doi:10.1002/pip.2766}). However, when compared to the CZTS growth for the tandem cell, significant morphological differences between the two CZTS layers are noticeable. The morphology comparison is presented in Figure \ref{fig:11}. The SEM top views of CZTS grown on the Si cell and on Mo/SLG, shown in Figure \ref{fig:11} (a) and (b), respectively, reveal a clear difference in grain size. The SEM cross-section of the tandem cell, in Figure \ref{fig:11} (c), shows that the CZTS exhibits a double layer structure, with a smaller grain size. In comparison, the CZTS grown with the same conditions on Mo-coated soda lime glass (SLG), shown in Figure \ref{fig:11} (d), has a single layer and larger grains. This indicates that the local conditions for CZTS growth are different in the two cases. In addition, CZTS photoluminescence (PL) measurements made on both the fully finished tandem and the thin CZTS cell, confirm that the thin CZTS cell has a significantly higher PL intensity (see supplementary Figure S12). We suggest that one possible reason for these differences is likely to be the contribution of Na diffusion from the glass, which is not available in Si. This possibility will be explored in future work.

The results of this work show that there is a margin for monolithically integrating a CZTS top cell on a full Si bottom cell using high temperature processing above 500 \degree C, without compromising the bottom cell. Given the constituting contaminant elements of CZTS (in particular Cu and S), we suggest that this study could be generalized to other thin film chalcogenide materials (many of which do contain Cu), and thereby open up the possibility of exploring new emerging wide band gap semiconductors as top cell alternatives.

\section{Conclusion}
We have assessed the potential of monolithically-integrated two-terminal tandem cells based on thin-film chalcogenides on Si, using CZTS and double-sided TOPCon Si as model system. We have investigated the use of a thin TiN barrier layer to protect the bottom Si cell from in-diffusion of metals and sulfur during the CZTS growth, and serve as interface recombination layer between the top and bottom cells at the same time. It was revealed that Cu contamination induced by CZTS growth on the Si bulk is significantly smaller than that from annealing of metallic Cu on Si. While traces of all CZTS elements (except for Sn) can be detected at the surface of c-Si after CZTS annealing, it was shown that the main contributor to the lifetime reduction in the bottom Si cell is Cu. Furthermore, it was shown that a TiN barrier layer as thin as 10 nm can effectively suppress the formation of Cu-related deep defects in Si. Based on these results, we presented a proof-of-concept monolithically integrated CZTS/Si tandem solar cell with an efficiency of 1.1\%  and a \textit{V}\textsubscript{oc} of 900 mV, which shows an additive \textit{V}\textsubscript{oc} effect. The i-\textit{V}\textsubscript{oc} of the silicon bottom cell was retained during the full fabrication of the CZTS cell when a 10 nm TiN barrier was used. It is suggested that the poor performance of the tandem cell is mainly due to limitations in the CZTS top cell, namely difficulty of reproducing high-quality CZTS absorbers on non-glass substrates, where Na is not available. The possibility of non-ohmic blocking behavior at the TiN interfaces is also mentioned.

By showing that a full TOPCon Si solar cell can be processed at temperatures well above 500  \degree C  in the presence of several critical contaminant elements – notably copper – without suffering from a severe degradation in lifetime and without forming deep defect levels, this work opens up the possibility of exploring other less known and future high bandgap compounds processed at high temperatures. This could allow for achieving high efficiency monolithically integrated tandem solar cells in the future.

\section{Acknowledgements}

A. Hajijafarassar (A.H.) and F. Martinho (F.M.) contributed equally to this work. This work was supported by a grant from the Innovation Fund Denmark (grant number 6154-00008A). F.M. would like to thank A. A. S. Lancia for the support on the J-V measurements.

\section{Data Availability}
All the data is available from the corresponding authors upon reasonable request.

\section*{References}

\bibliography{mybibfile}

\end{document}


\begin{frontmatter}

\title{Monolithic Thin-Film Chalcogenide-Silicon Tandem Solar Cells Enabled by a Diffusion Barrier}


\author[mymainaddress]{Alireza Hajijafarassar\corref{mycorrespondingauthor}\fnref{myfootnote}}
\ead{alhaj@dtu.dk}

\author[mysecondaryaddress]{Filipe Martinho\corref{mycorrespondingauthor}\fnref{myfootnote}}
\ead{filim@fotonik.dtu.dk}
\fntext[myfootnote]{These authors contributed equally.}
\cortext[mycorrespondingauthor]{Corresponding authors}

\author[mythirdaddress]{Fredrik Stulen}

\author[mythirdaddress]{Sigbjørn Grini}

\author[mymainaddress]{Simón López-Mariño}

\author[mysecondaryaddress]{Moises Espíndola-Rodríguez}

\author[myforthaddress]{Max Döbeli}

\author[mysecondaryaddress]{Stela Canulescu}

\author[mymainaddress]{Eugen Stamate}

\author[mysecondaryaddress]{Mungunshagai Gansukh}

\author[mysecondaryaddress]{Sara Engberg}

\author[myfifthaddress]{Andrea Crovetto}

\author[mythirdaddress]{Lasse Vines}

\author[mysecondaryaddress]{Jørgen Schou}

\author[mymainaddress]{Ole Hansen}

\address[mymainaddress]{DTU Nanolab, Technical University of Denmark, DK-2800 Kgs. Lyngby, Denmark}

\address[mysecondaryaddress]{Department of Photonics Engineering, Technical University of Denmark, DK-4000 Roskilde, Denmark}

\address[mythirdaddress]{Department of Physics, University of Oslo, 0371 Oslo, Norway}

\address[myforthaddress]{Ion Beam Physics, ETH Zurich, CH-8093 Zurich, Switzerland}

\address[myfifthaddress]{DTU Physics, Technical University of Denmark, DK-2800 Kgs. Lyngby, Denmark}

\end{frontmatter}

\linenumbers

\section{Results and Discussion (Supplementary)}
\begin{figure}
  \centering
  \includegraphics[width=0.6\linewidth]{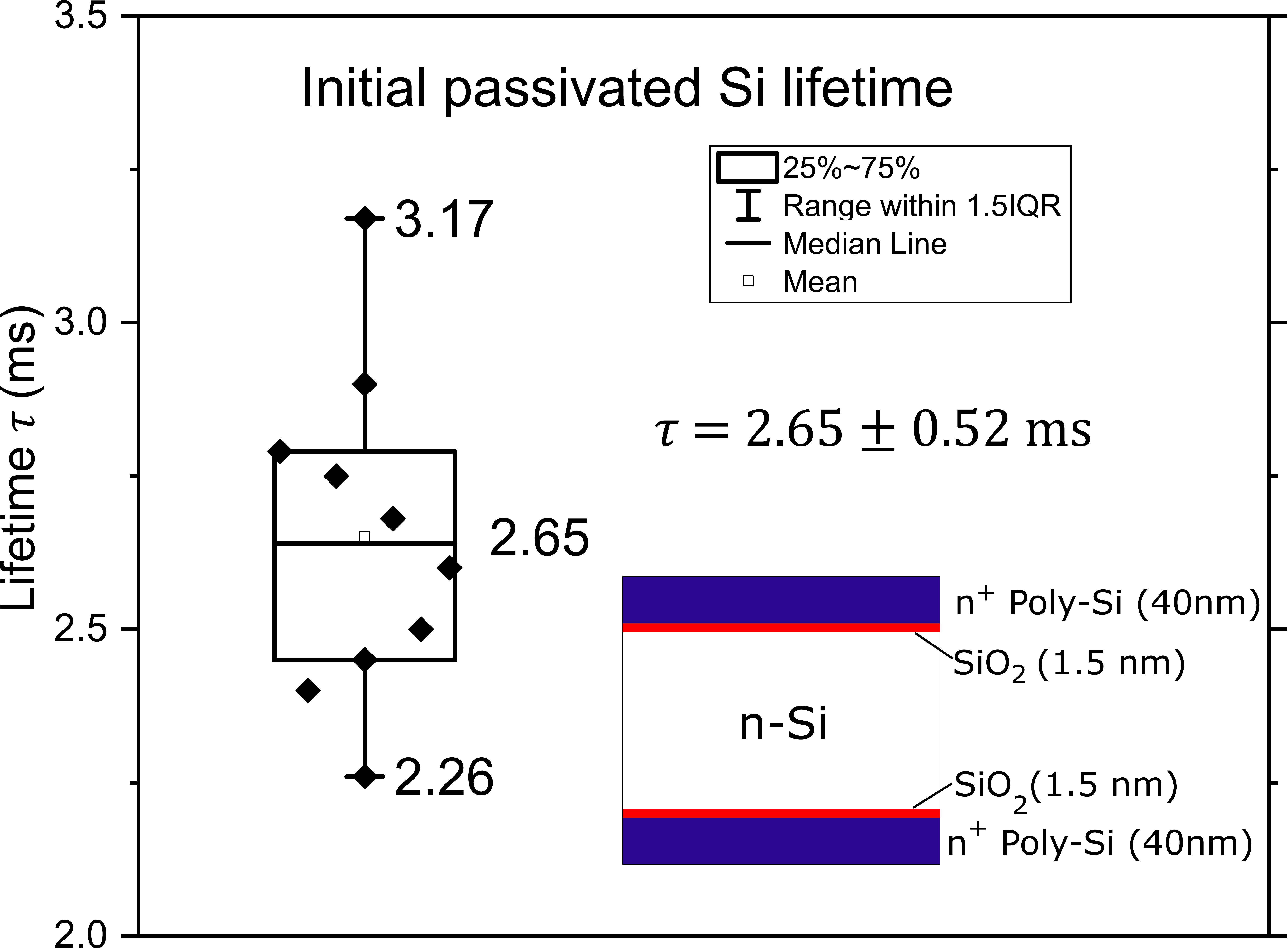}
  \caption{Statistical lifetime distribution of the set of 10 symmetrically n-passivated wafers. A scheme of the passivated structure is shown as inset).}
    \label{fig:s1}
\end{figure}

The statistics for the set of 10 symmetrically n-passivated wafers is shown in Figure \ref{fig:s1}. The representative lifetime was taken as the mean value, and the uncertainty as the maximum deviation. The passivation quality is considered to be high, as the corresponding implied V\textsubscript{oc} of the silicon solar cell exceeds 700 mV. It is important to have a good quality passivation when conducting lifetime studies, because the effective minority carrier lifetime is always a contribution from bulk recombination and surface recombination. A high quality surface passivation minimizes the contribution of the surface recombination term, and allows to draw conclusions about changes in the bulk during processing. Throughout this study, it was observed that samples with low quality passivation (or weak resistance to depassivation) would yield different lifetime results due to the depassivation of the surface, making it difficult to study specific changes in the bulk due to contamination. By ensuring a high quality passivation across the wafer set, we have more confidence that the lifetime results are directly comparable and give information about changes in the Si bulk.

\begin{figure}
\centering
\begin{subfigure}[b]{.47\textwidth}
  \centering
  \includegraphics[width=\linewidth]{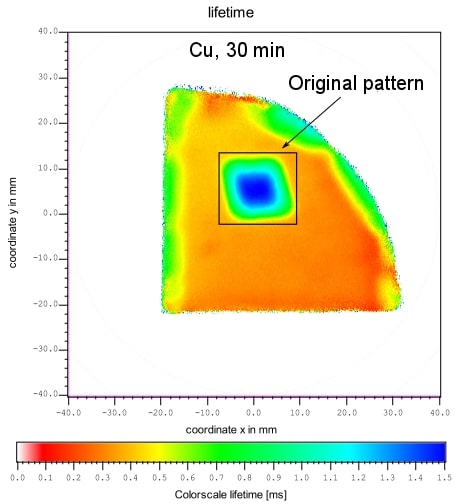}
  \caption{}
  \label{fig:2sa}
\end{subfigure}
\begin{subfigure}[b]{.47\textwidth}
  \centering
  \includegraphics[width=\linewidth]{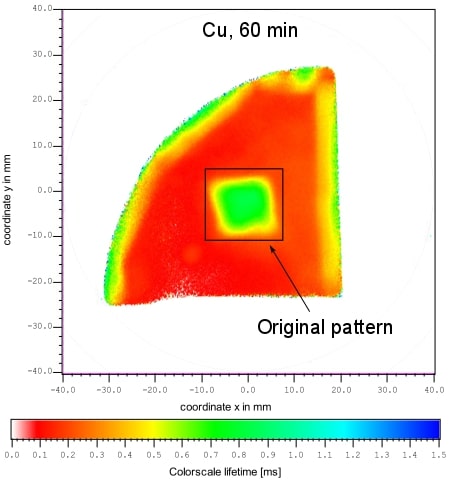}
  \caption{}
  \label{fig:2sb}
\end{subfigure}
\caption{Variation of the Cu diffusion reference of the main text with patterning included. (a) After just 30 minutes of annealing in N\textsubscript{2}, the edges of the pattern are already showing decreased lifetime, highlighting the effects of lateral diffusion of Cu. The region not covered with Cu, defined by the pattern, retains a high lifetime, similar to the as-passivated value, meaning that the surface does not depassivate during annealing; (b) After 60 minutes of annealing in N\textsubscript{2}, some Cu already diffused into the patterned area, as indicated by the further degradation in lifetime.}
\label{fig:s2}
\end{figure}

In the metallic \textit{Cu reference} lifetime test, to prove unequivocally that it is the metallic Cu that is responsible for the deterioration in lifetime (and not, for example, the high temperature annealing of its own and possible effects on the passivation), we conducted a similar metallic Cu diffusion test as in the main text, but patterned a large area near the center of the wafer, where no Cu was deposited. The results are shown in Figure \ref{fig:s2}. After annealing, the results clearly show that the Cu-free area exhibits a lifetime close to the original as-passivated values ($>$ 1.5 ms), whereas the areas exposed to Cu show a large degradation in lifetime. Moreover, by keeping track of changes the annealing time, the effect of lateral Cu diffusion into the patterned area is evident, highlight the high diffusivity of Cu in Si.

\begin{figure}
  \centering
  \includegraphics[width=0.6\linewidth]{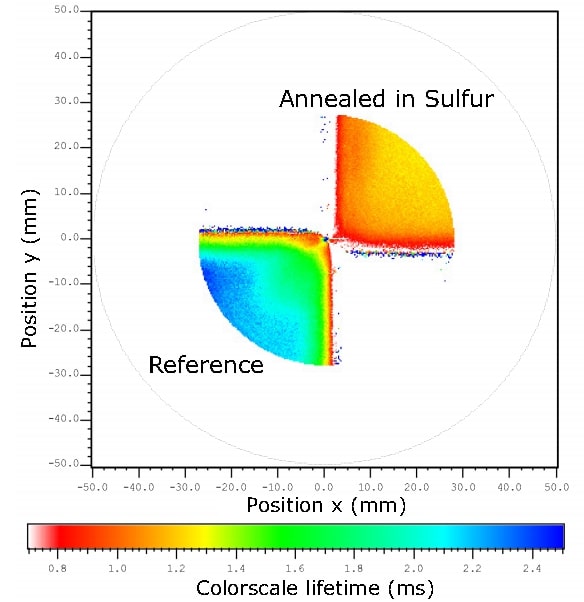}
  \caption{Lifetime mapping showing the effeccts of sulfur annealing for the 25 nm TiN case at 525  \degree C, compared to a non-annealed quarter from the same wafer. It clearly shows the detrimental effect of annealing the Si wafer in a sulfur atmosphere.}
    \label{fig:s3}
\end{figure}

The \textit{Sulfur reference} test allowed us to decouple the effect of the S atmosphere during the CZTS annealing from the effects of the CZTS itself. In Figure \ref{fig:s3} a direct comparison of an as-passivated quarter and a quarter annealed in an S atmosphere is shown. Both quarters are from the same initial wafer. The result indicates that S is detrimental for the lifetime of Si, which explains the lifetime results shown in the main text. Interestingly, the final after annealing lifetime is nearly the same for S annealing and for CZTS annealing. Considering the SIMS, RBS, and DLTS results from the main text, this suggests that the mechanism for lifetime degradation is different in these two cases: in the sulfur annealing it is sulfur contamination which is degrading the lifetime, whereas in CZTS annealing the Cu contamination seems to be more relevant. Still, in both cases the end lifetimes are significantly higher than in the metallic Cu case.

\begin{figure}
  \centering
  \includegraphics[width=0.6\linewidth]{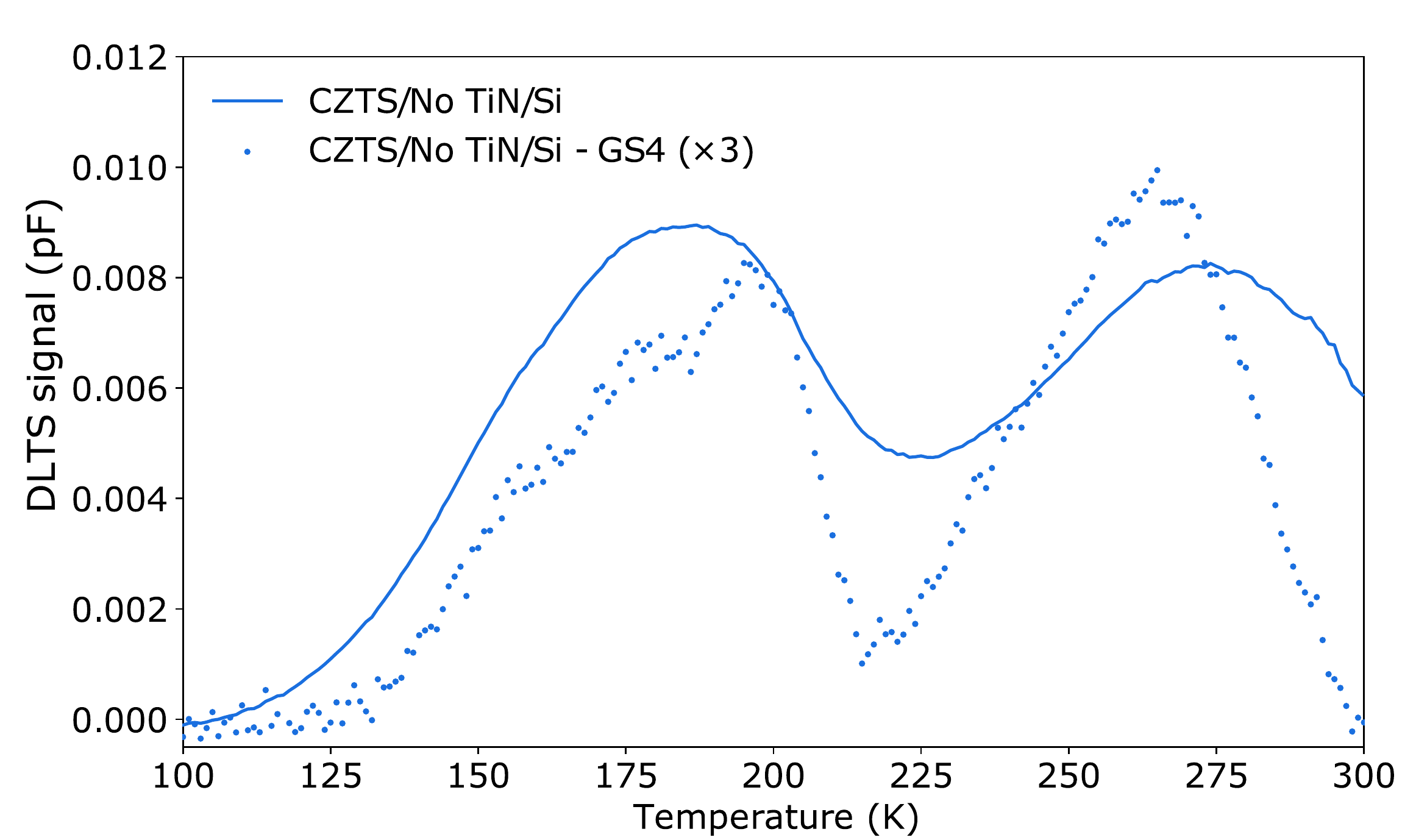}
  \caption{Comparison of the main DLTS signal (lock-in) with the 4-term Gaver–Stehfest (GS4) signal. The comparison shows that the low temperature (below 200 K) feature is a contribution of several peaks, whereas the high temperature feature (around 270 K) is a single peak. Here, the rate window used was 640 ms (window 6).}
    \label{fig:s4}
\end{figure}

In the main text, a claim was made that the interpretation of the low temperature DLTS peak near 175 K was ambiguous due to the low temperature tail and broadening. To gain further insight on this, a 4-term Gaver–Stehfest algorithm (GS4) was used for signal extraction. The advantage of this algorithm is that it significantly increases the detection resolution, at the cost of a lower signal-to-noise ratio \cite{doi:10.1063/1.366269}. The results are shown in Figure \ref{fig:s4}. Using GS4, several individual peaks can now be resolved. This is an indication of either overlapping or extended defects in the Si bulk, and justifies why no activation energies were extracted for this low temperature feature. The extraction was only done for the higher temperature peak (near 270 K).

\begin{figure}
\centering
\begin{subfigure}[b]{.47\textwidth}
  \centering
  \includegraphics[width=\linewidth]{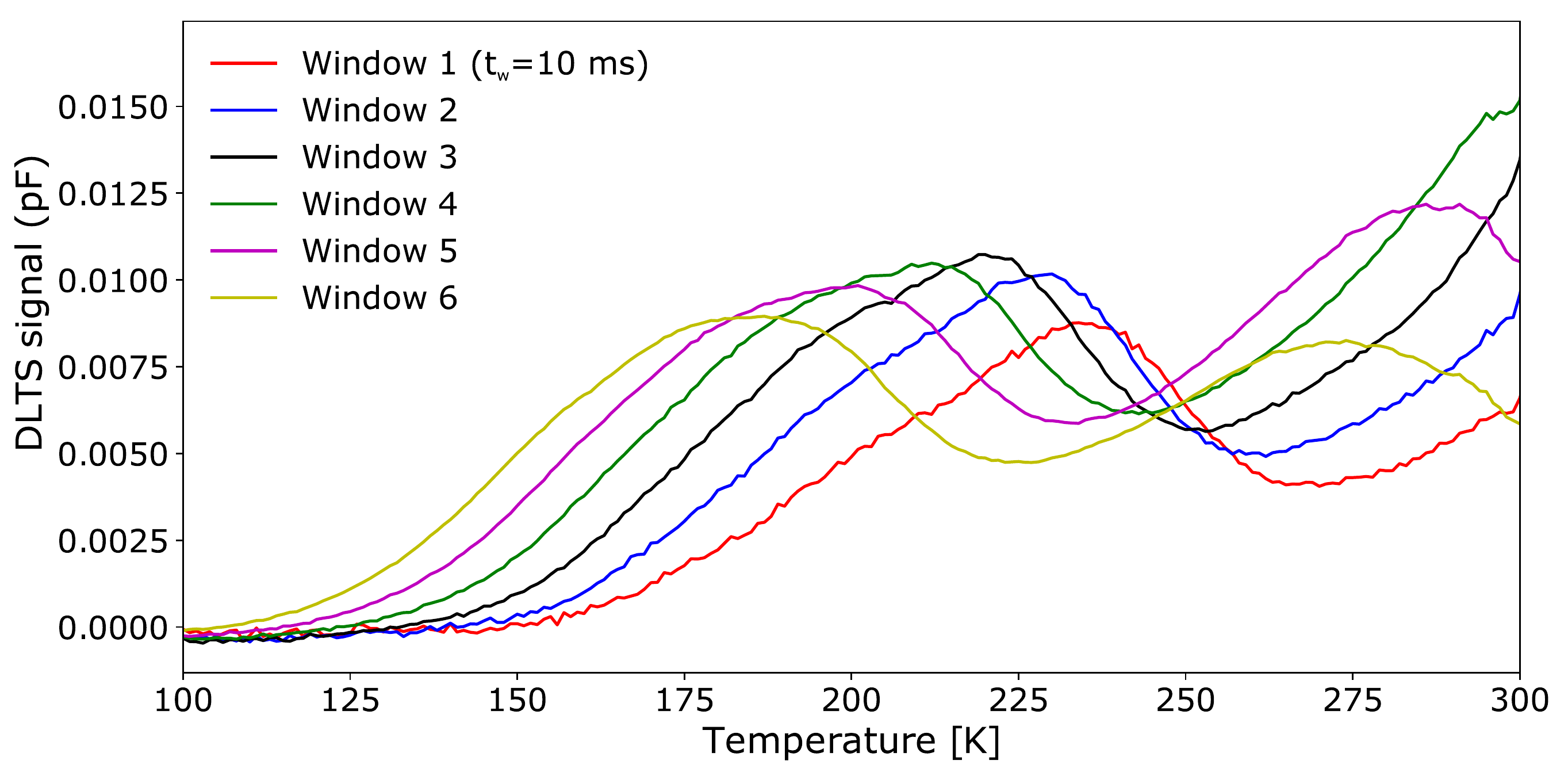}
  \caption{}
  \label{fig:5sa}
\end{subfigure}
\begin{subfigure}[b]{.41\textwidth}
  \centering
  \includegraphics[width=\linewidth]{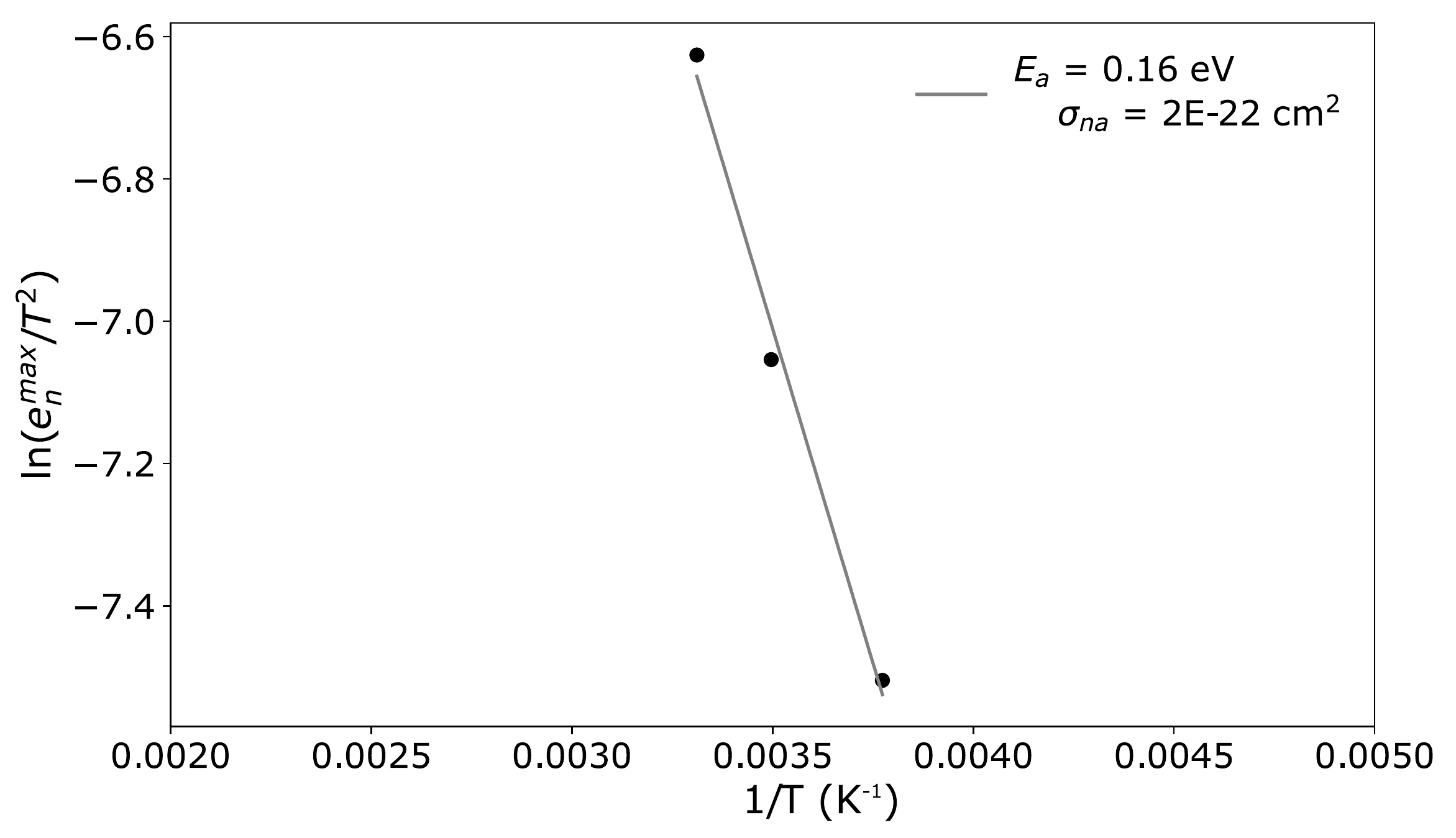}
  \caption{}
  \label{fig:5sb}
\end{subfigure}
\caption{DLTS spectra with time window variation (with lengths 2$^i\times$10 ms, $i$ = 1, $\ldots$ , 6); (b) Arrhenius plot corresponding to the high temperature peak of (a). Here, e\textsubscript{n}\textsuperscript{max} is the electron emission rate, E\textsubscript{a} is the defect activation energy, and  $\sigma$\textsubscript{na} is the capture cross-section. Only 3 peaks fall within the experimental temperature range.}
\label{fig:s5}
\end{figure}

Having identified the high temperature feature as a single peak, a rate window variation was performed to extract the corresponding Arrhenius plot from the peak shift. These results are shown in Figure \ref{fig:s5} (a) and (b). Certain peaks shifted outside the measurable temperature range, meaning that only 3 peaks could be used to make the Arrhenius plot. As such, the extracted parameters have a large uncertainty. Nevertheless, the values reasonably match well-known Cu defects in Si, as mentioned in the main text.

\begin{figure}
  \centering
  \includegraphics[width=0.6\linewidth]{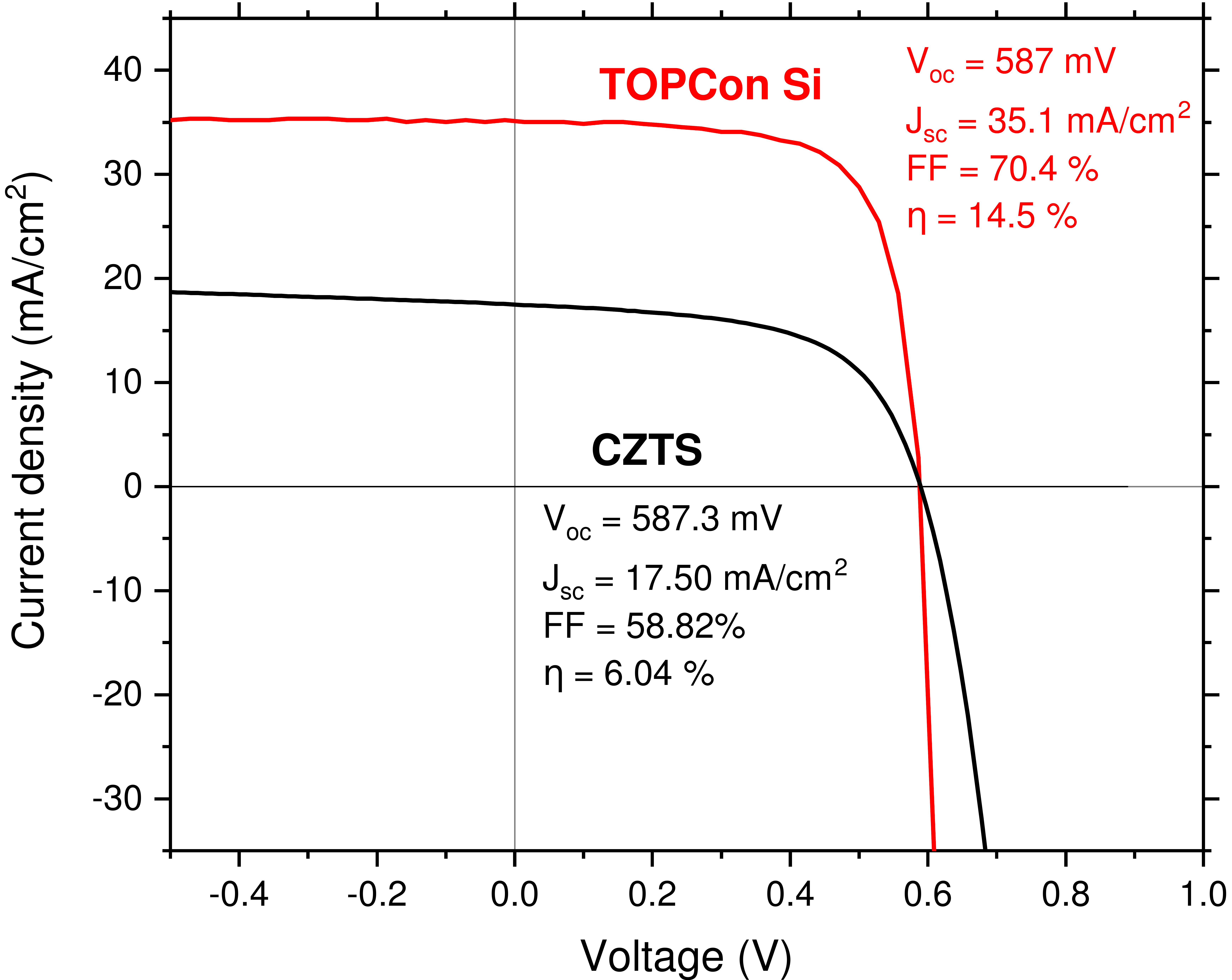}
  \caption{The best single junction J-V characteristics of our processing line at the time of fabrication of the CZTS/Si tandem cell. In the tandem cell, the V\textsubscript{oc} reached close to 900 mV, whereas each individual cell only shows around 600 mV.}
    \label{fig:s6}
\end{figure}

In the text, it was claimed that the tandem exhibited a Voc of 900 mV, and higher than each individual single junction solar cell. Figure \ref{fig:s6} shows the best J-V curves of both single junction solar cells, at the time the CZTS/Si tandem was fabricated. Although the values are currently under optimization, it can be seen that in both cases a Voc around 600 mV is expected using the same parameters as used for the production of the tandem cell (when applicable).

\begin{figure}
  \centering
  \includegraphics[width=0.6\linewidth]{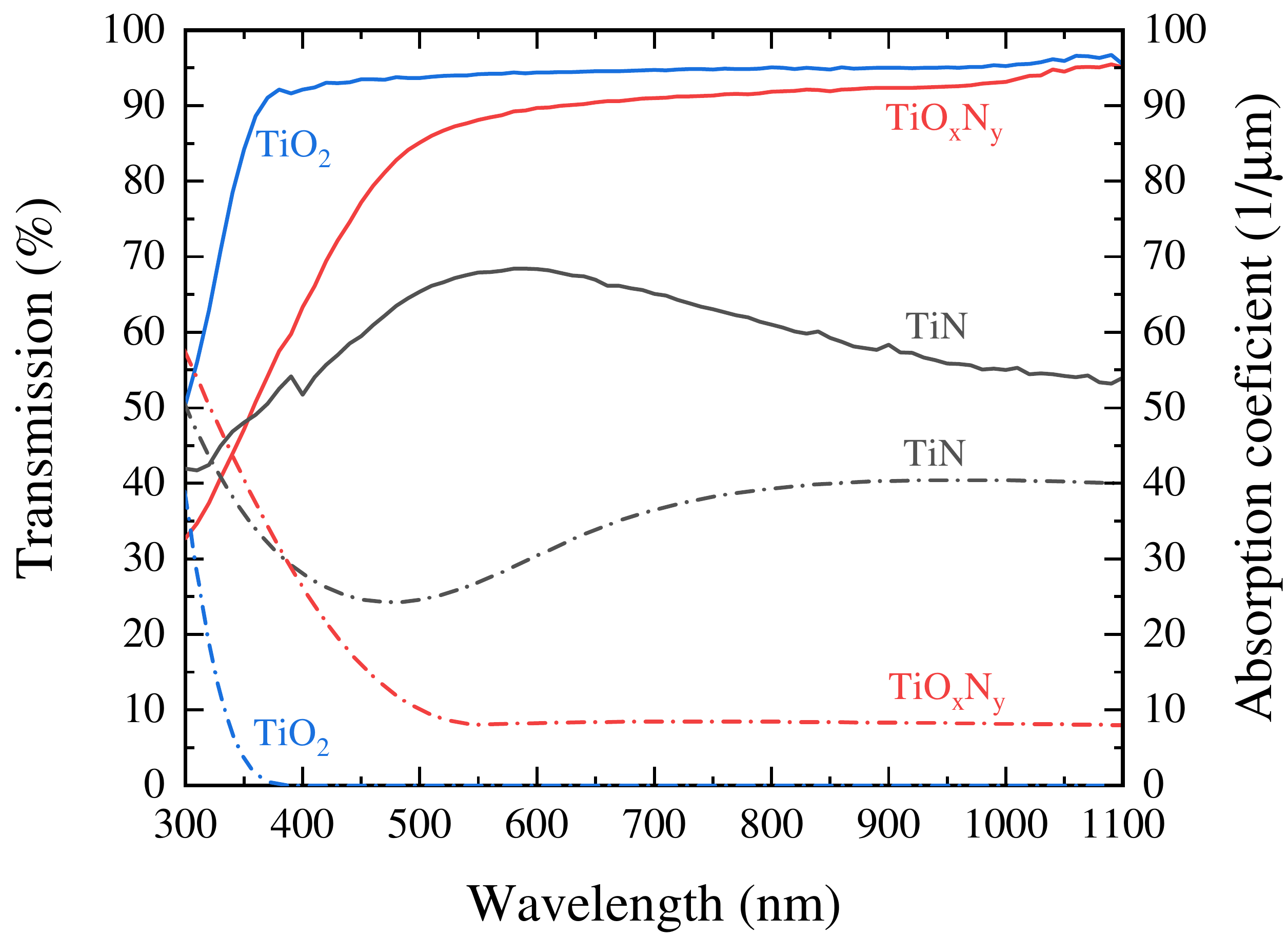}
  \caption{Transmission spectra (solid lines left y-axis) and aborsption coeficient (dashed lines right y-axis) of 10nm TiN, 10 nm TiO\textsubscript{x}N\textsubscript{y}, and 10 nm TiO\textsubscript{2} films deposited on fused-silica (SiO\textsubscript{2}) substrates.}
    \label{fig:s12}
\end{figure}

As mentioned in the main text, the transparency of the TiN layer can be improved by incorporating oxygen into the films, albeit at the cost of increased resistivity. In this work, the oxygen content was controlled by the amount of residual oxygen available in the ALD reactor during the deposition. The transmission spectrum as well as the absorption coeficient ($\alpha$) of the 10 nm TiO\textsubscript{x}N\textsubscript{y}, used in the tandem cell, is compared with 10 nm TiN and TiO2 layers in Figure \ref{fig:s12}. As can be seen in Figure \ref{fig:s12}, a high transmission of around 90 \% in the near infrared region was achieved for the TiO\textsubscript{x}N\textsubscript{y} layer, however, the resisitivity of the film increased by several orders of magnitude from 0.5 - 1 m$\Omega$.cm for TiN, to around 40 $\Omega$.cm for TiO\textsubscript{x}N\textsubscript{y}. 

\begin{figure}
\centering
\begin{subfigure}[b]{.464\textwidth}
  \centering
  \includegraphics[width=\linewidth]{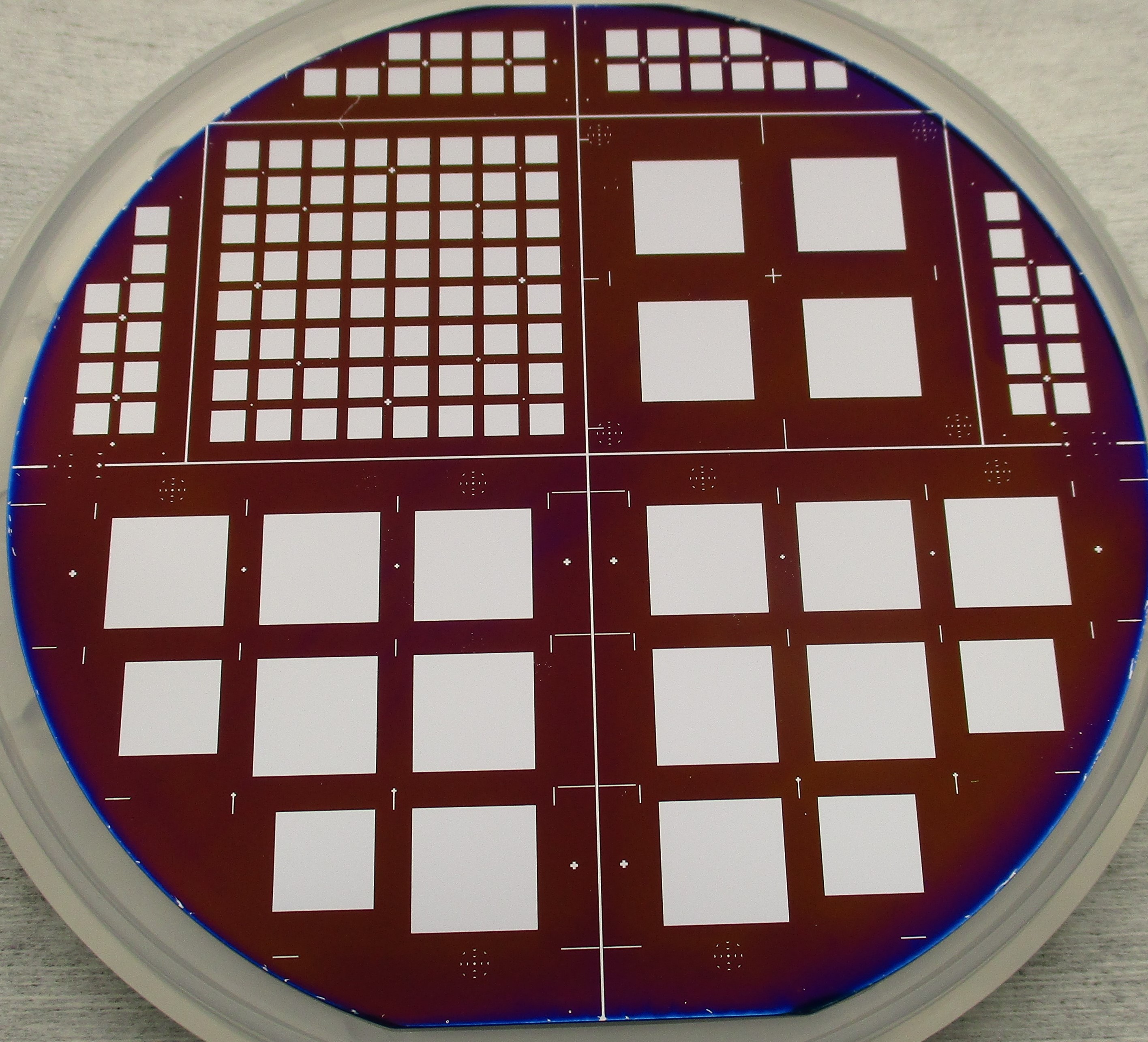}
  \caption{p+PolySi/TO}
  \label{fig:s13a}
\end{subfigure}
\begin{subfigure}[b]{.47\textwidth}
  \centering
  \includegraphics[width=\linewidth]{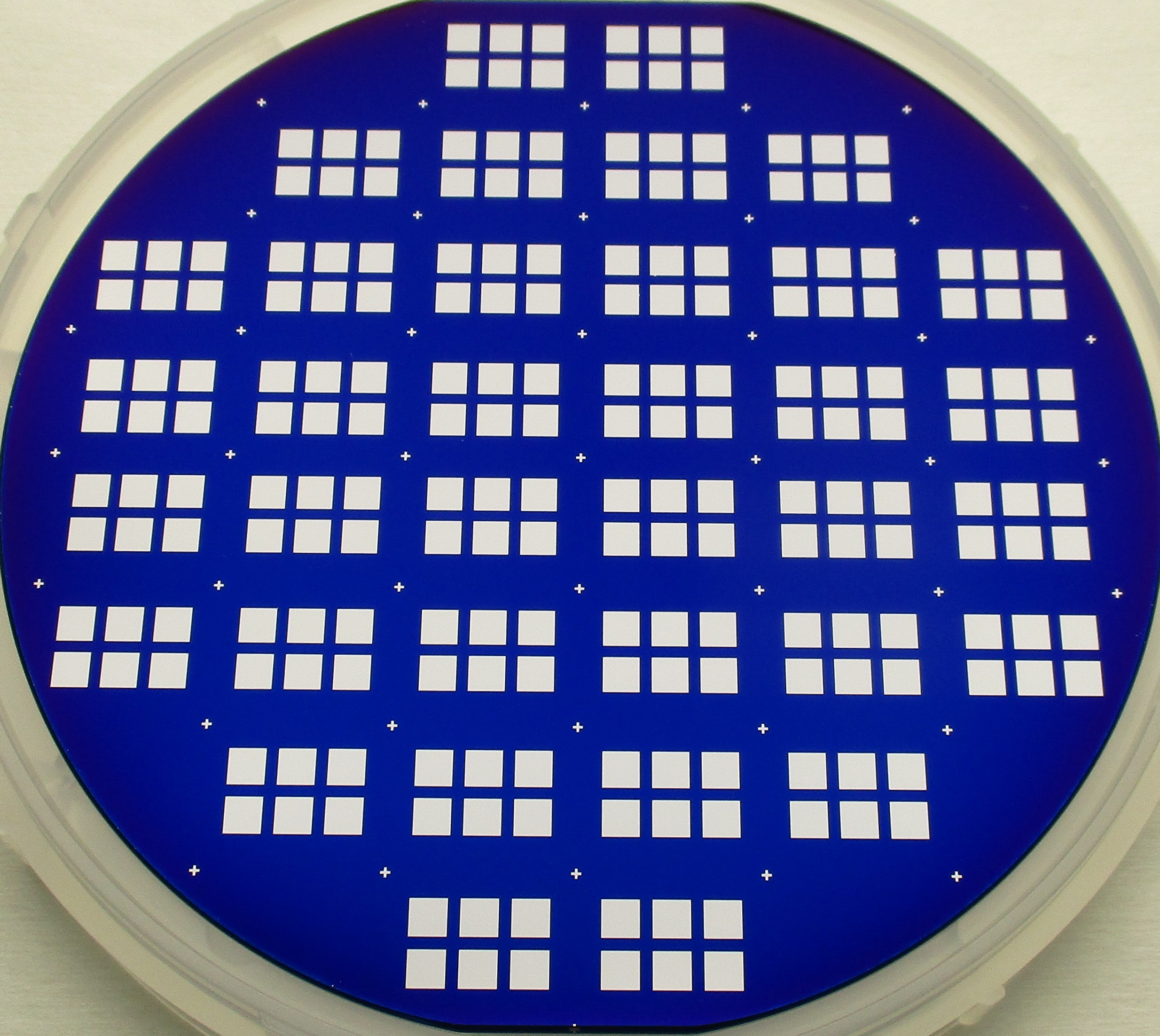}
  \caption{p+PolySi/TO}
  \label{fig:s13b}
\end{subfigure}
\begin{subfigure}[b]{.47\textwidth}
  \centering
  \includegraphics[width=\linewidth]{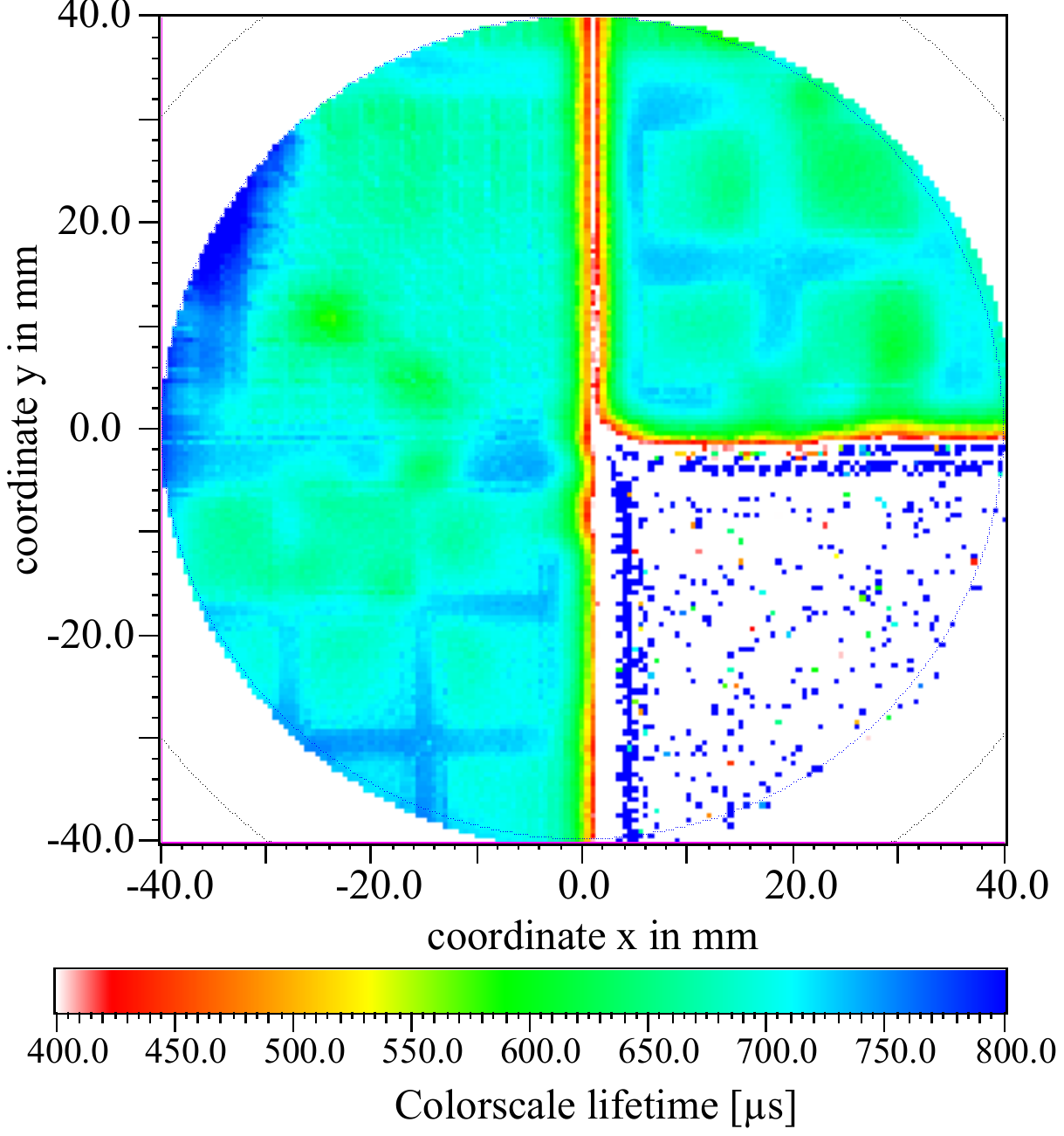}
  \caption{TiN on n+PolySi/TO}
  \label{fig:s13b}
\end{subfigure}
\begin{subfigure}[b]{.47\textwidth}
  \centering
  \includegraphics[width=\linewidth]{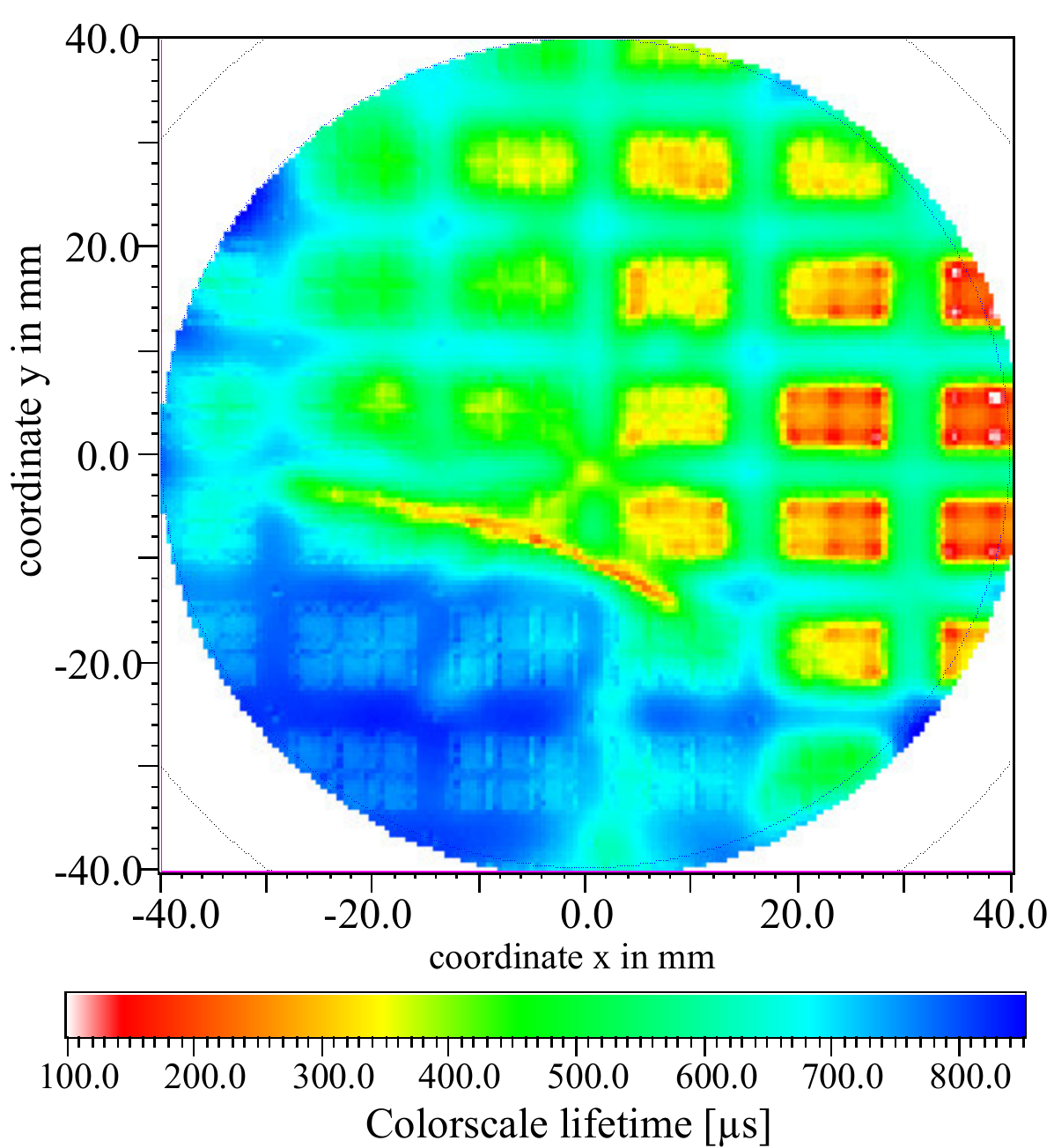}
  \caption{TiN on p+PolySi/TO}
  \label{fig:s13b}
\end{subfigure}
\caption{Effect of the TiN deposition on the passivisation quality of n+polySi (left) and p+polySi (right) TOPCon structures. Patterned SiN pictures ((a) and (b)) and the corresponding lifetime maps ((c) and (d)) on n+polySi/TO and p+polySi/TO structures.}
\label{fig:s13}
\end{figure}

By monitoring the minority carrier lifetime of the samples throughout different processing steps, we  observed that the TiN deposition can have a negative effect on the lifetime of the samples. In section 3.1 of the main text, it was argued that for the samples in the lifetime series a mild Fe contamination could be occurring during the ALD step, and in the tandem cell sample an additional partial loss in the hydrogen passivation could be happening as well, as the sample had an extra hydrogenation step using SiN:H. To further evaluate the TiN effect, special samples were prepared. In these samples, the front (n+PolySi) or rear (p+PolySi) surface of the device precursor wafers were coated by a 75 nm SiN:H layer. The SiN:H layer was then patterned by photo-lithography and wet-etching in buffered HF to expose the underlying n+polySi/ p+polySi layer. Photos of such samples are shown in Figure \ref{fig:s13} (a) and (b) as examples. It has been confirmed by lifetime maps (before and after patterning, not shown here) that the patterning process has no significant effect on the minority carrier lifetime of the wafers. Moreover, these lifetime mappings have been done with the non-patterned side facing upward (which had a blanket SiN:H layer) to exclude any possible optical effects of the pattern on the measurements. Subsequently, 10 nm TiO\textsubscript{x}N\textsubscript{y} was deposited on the patterned surface, and the minority carrier lifetime was mapped over the entire wafer. The obtained lifetime maps are shown in Figure \ref{fig:s13} (c) and (d). As it can be noted from the lifetime maps, the exposed areas exhibit lower minority carrier lifetime in both cases, however the p+PolySi sample seems to have more severe degradation compared to the n+PolySi counterpart. A possible explanation for this loss can be attributed to a partial loss of hydrogen passivation (provided by the preceding hydrogenation process with SiN:H) in the exposed areas as a result of the high temperature ALD process. In the non-exposed areas, this loss is not expected as the hydrogen is provided by the SiN:H layer, which contains high amount of atomic hydrogen as impurity. If that is the case, a post-TiN hydrogenation process, e.g., by annealing the samples in a hydrogen rich-atmosphere, such as forming gas or remote hydrogen plasma, could help to recover the initial lifetime. Regarding the possibility of Fe contamination, it has been seen that placing the samples on a clean dummy wafer can mitigate the degradation (not shown here). A clean dummy wafer was in fact used for the tandem cell fabrication (in section 3.4 of the main text), but a degradation in lifetime was still seen after the TiN step, suggesting that the hydrogenation loss might also be playing a role. Note, however, that in the case of a mild contamination during the TiN deposition, the same protective effect of SiN shown in Figure S8 would still occur, as it would block the diffusion of contaminants into Si. For this reason, we are still not able to single out the reason for the degradation occurring during the TiN step, and this requires additional investigation in the future.

\begin{figure}
  \centering
  \includegraphics[width=0.6\linewidth]{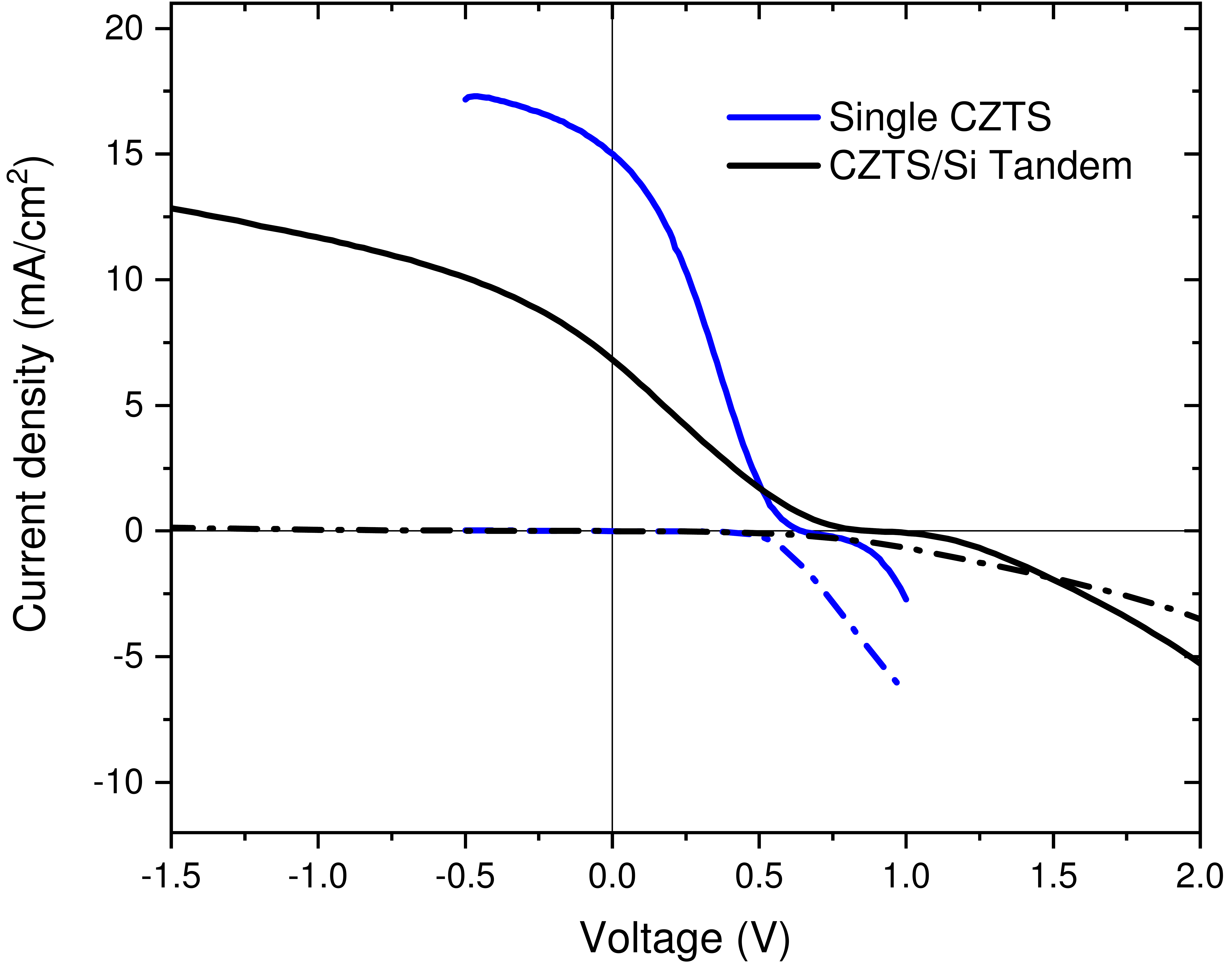}
  \caption{Comparison of a rollover effect experimentally found during the tuning of the CZTS baseline process and the roll-over effect on the tandem cell. While this problem was resolved in our single junction CZTS baseline, the reasons for this behavior in the tandem cell are currently under investigation.}
    \label{fig:s7}
\end{figure}

We have seen throughout our processing that blocking behaviors can occur also in our single junction CZTS cells, causing a roll-over effect on the J-V curve. We compare this to the tandem cell in Figure \ref{fig:s7}. In a tandem cell, we speculate that this could be related to the lack of alkali elements in the CZTS layer or to non-ohmic interfaces at the TiN barrier layer. In the case of the single CZTS cell, the roll-over was caused by a non-ideal CZTS composition, and the effect was completely removed by tuning the CZTS composition. In the tandem cell, the same ideal CZTS composition was used. The exact cause of the rollover behavior in the tandem cell is still under investigation.

\begin{figure}
  \centering
  \includegraphics[width=0.6\linewidth]{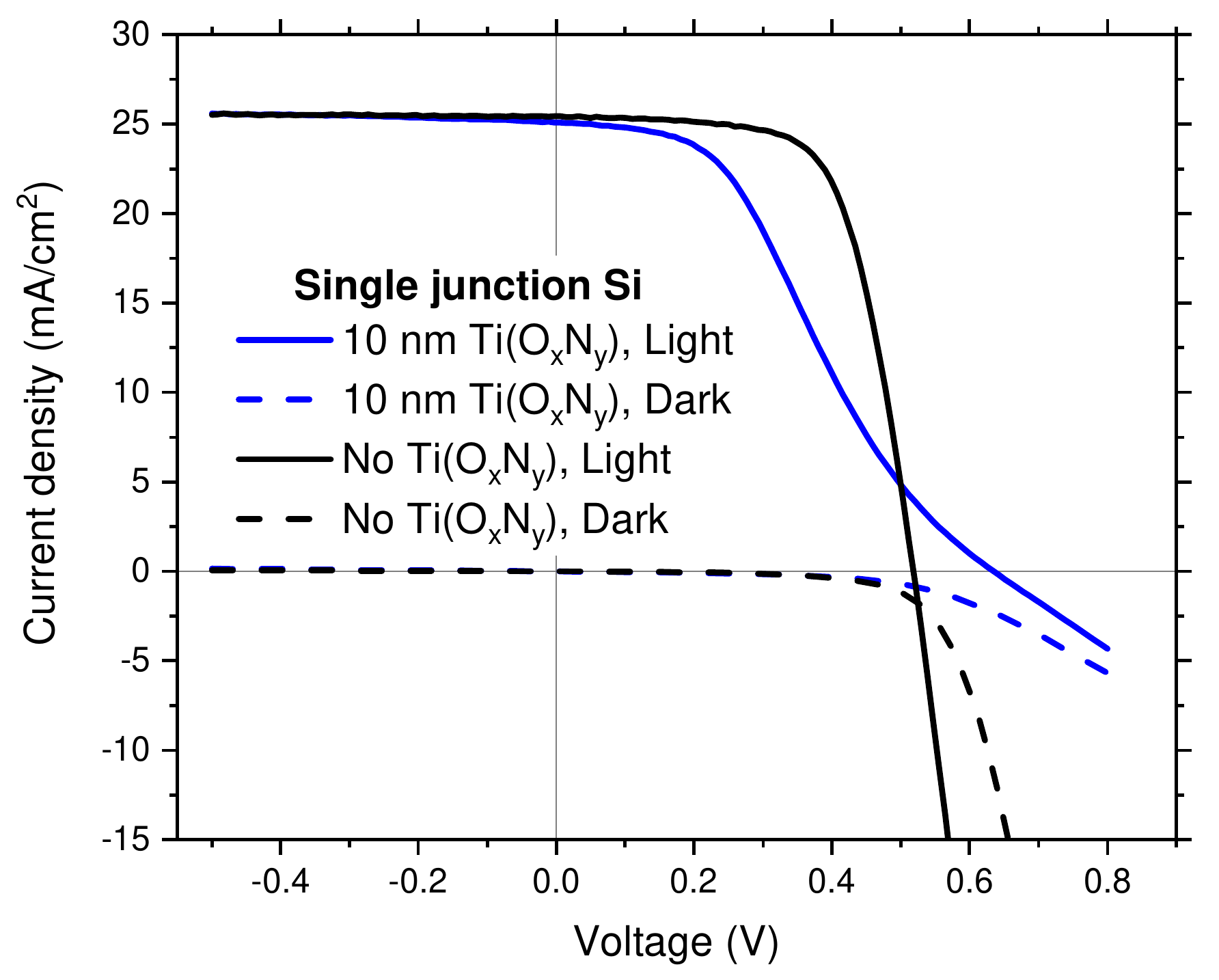}
  \caption{Comparison of a single junction Si solar cell with and without a TiN layer on the top. The presence of a 10 nm TiN layer causes a roll-over effect, distorting the J-V curve.}
    \label{fig:s8}
\end{figure}

Additionally, we have found evidence that part of the roll-over effect could be occurring on the Si bottom cell alone, due to the presence of TiO\textsubscript{x}N\textsubscript{y} between the n+polySi and the TCO layers. This is shown in the J-V curve of Figure \ref{fig:s8}. The roll-over is not as significant as in the tandem cell, suggesting that this is likely not the only contributor to the overall distortion of the tandem J-V curve. As mentioned in the main text, this effect can be controlled by tuning the amount of oxygen allowed in the TiN film. 

\begin{figure}
  \centering
  \includegraphics[width=0.6\linewidth]{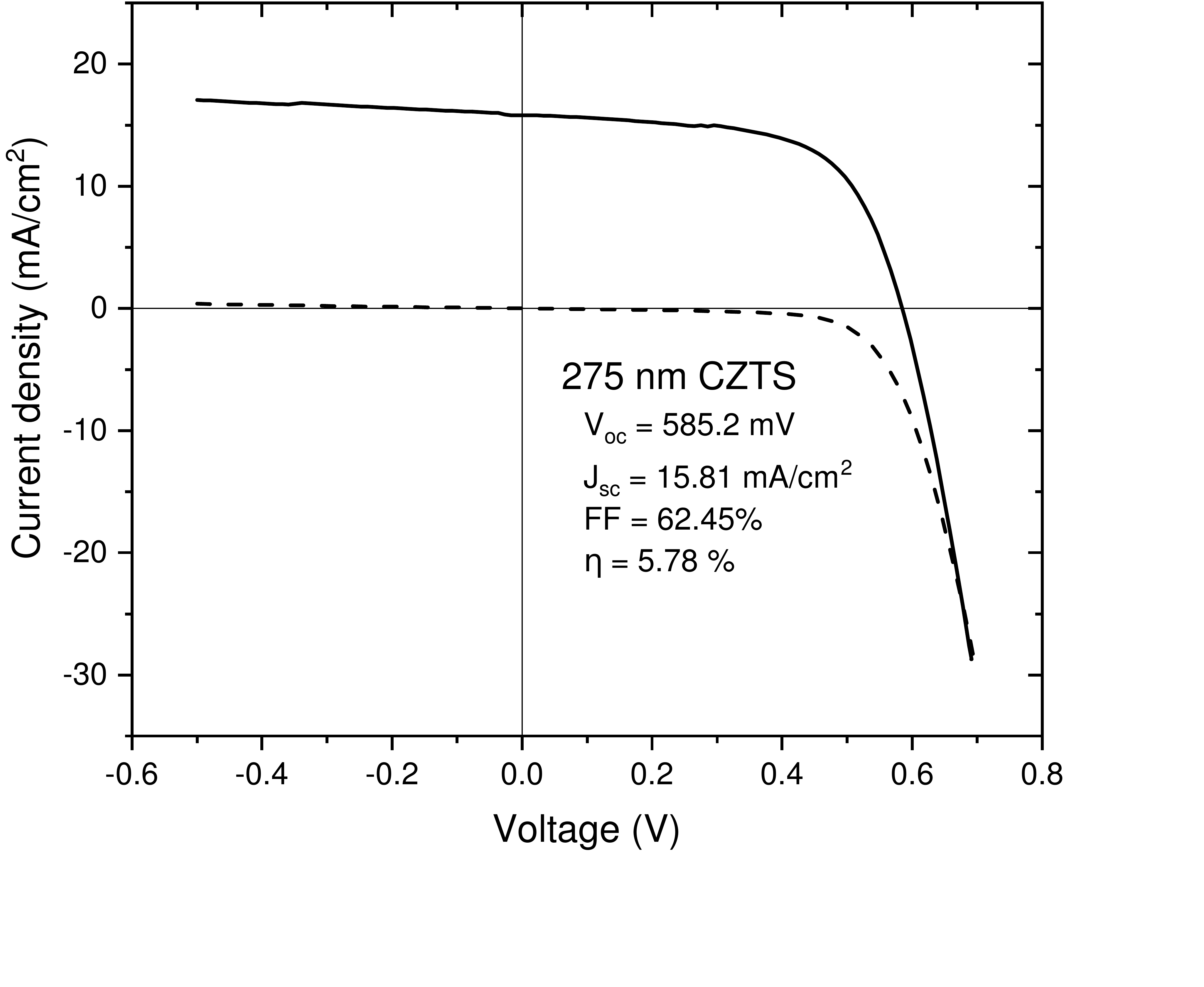}
  \caption{J-V characteristic of the single junction thin CZTS solar cell. Despite a reduced thickness of only 275 nm, the current reaches close to 16 mA/cm\textsuperscript{2}.}
    \label{fig:s9}
\end{figure}

To evaluate the performance of the top cell and the current matching condition, a CZTS single junction solar cell was prepared, with a thickness similar to that of the tandem cell, around 275 nm. The J-V characteristic of this $``$thin single thin CZTS cell$"$  is plotted in Figure \ref{fig:s9}. Despite its reduced thickness, this solar cell reaches almost 16 mA/cm\textsuperscript{2} and has an efficiency of 5.8$\%$ , comparable to the best CZTS cells produced at the time of the tandem fabrication. However, the tandem results do not show this behavior, as can be seen by the severe blocking occurring in the first quadrant of the J-V curves. This suggests that the top CZTS cell in the tandem is not performing as well as the thin single junction CZTS. The reasons for this are currently under investigation.

\begin{figure}
  \centering
  \includegraphics[width=0.6\linewidth]{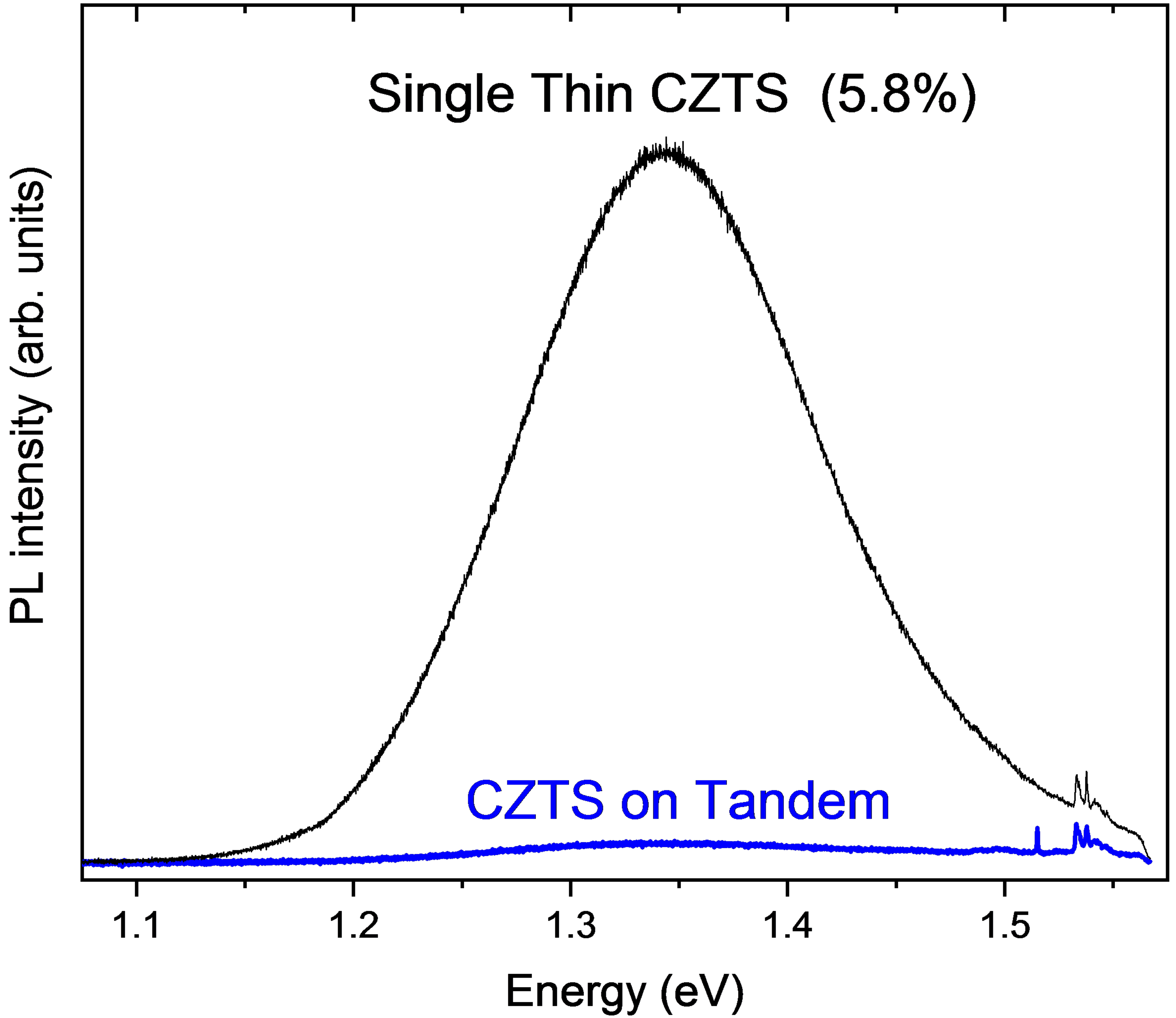}
  \caption{Photoluminescence comparison of CZTS grown on 10 nm TiN in the tandem device (blue) and on Mo in a 5.8$\%$  efficient thin CZTS cell (black). The wavelength of the excitation laser was 785 nm.}
    \label{fig:s10}
\end{figure}

A further indication of the top CZTS cell limiting the tandem performance was obtained using room-temperature photoluminescence (PL) on complete devices. As Figure \ref{fig:s10} shows, the single thin CZTS cell has a PL yield 2 orders of magnitude higher than the CZTS on the tandem cell, despite having been produced under the same conditions (including all the buffer layers). This naturally comes from the difference in grain size, as shown by the SEM images in the main text, but it is also debated that Na and other alkali metals can lead to a reduction in non-radiative recombination and increase in device performance \cite{doi:10.1063/1.4916635}.

\begin{figure}
\centering
\begin{subfigure}[b]{.45\textwidth}
  \centering
  \includegraphics[width=\linewidth]{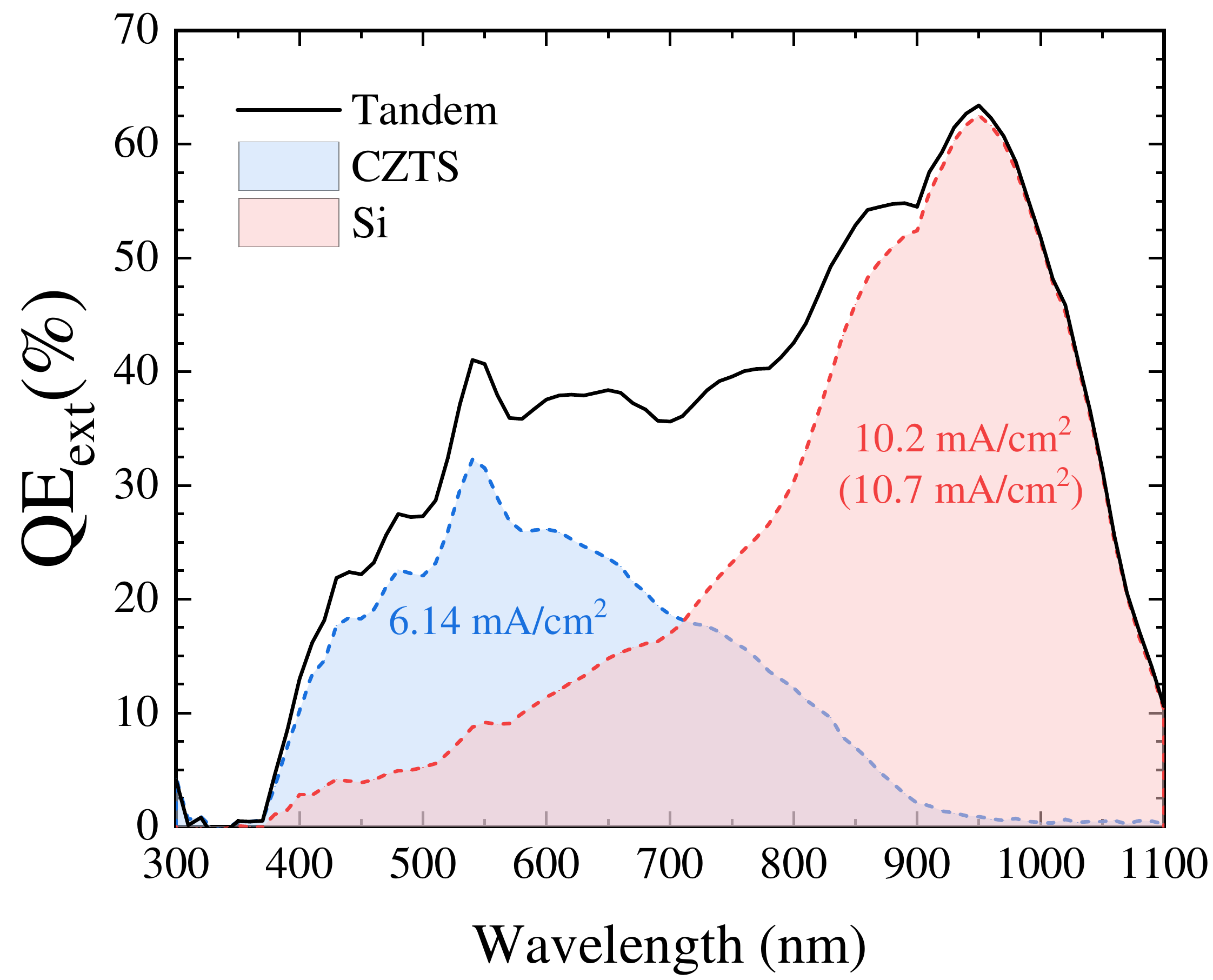}
  \caption{}
  \label{fig:1a}
\end{subfigure}
\begin{subfigure}[b]{.45\textwidth}
  \centering
  \includegraphics[width=\linewidth]{images/image20.pdf}
  \caption{}
  \label{fig:1b}
\end{subfigure}
\caption{EQE Improvement of the top cell upon hot plate annealing in air at 250 \degree C: (a) before hot plate annealing; (b) after hotplate annealing.}
\label{fig:s11}
\end{figure}

It is mentioned in the experimental methods that the full tandem cell is annealed on the hotplate in air at 250 \degree C. This step is used to improve the properties of the CZTS/CdS heterojunction. The effects of this annealing step are shown in Figure \ref{fig:s11}.

\section*{References}

\bibliography{mybibfile}